\documentclass[oldversion]{aa}
\usepackage{graphicx}
\usepackage{txfonts}
\usepackage{natbib}
\bibpunct{(}{)}{;}{a}{}{,} % to follow the A&A style

\newcommand{\sqdeg}{\mbox{deg$^{2}$}}

\newcommand{\yk}{\mbox{$Y\!-\!K$}}

\newcommand{\dmo}{\mbox{$(m\!-\!M)_{0}$}}
\newcommand{\ebv}{\mbox{$E_{B\!-\!V}$}}
\newcommand{\eyj}{\mbox{$E_{Y\!-\!J}$}}
\newcommand{\eyk}{\mbox{$E_{Y\!-\!K}$}}

\newcommand{\mh}{\mbox{\rm [{\rm M}/{\rm H}]}}
\newcommand{\Msun}{\mbox{$M_{\odot}$}}

\newcommand{\comment}[1]{}
\newcommand{\beq}{\begin{equation}}
\newcommand{\eeq}{\end{equation}}
\newcommand{\beqa}{\begin{eqnarray}}
\newcommand{\eeqa}{\end{eqnarray}}
\begin{document}
\title{On the Recovery of the Star Formation History of the LMC
from the VISTA Survey of the Magellanic System} 

\author{Leandro Kerber\inst{1,2}
\and L\'eo Girardi\inst{1}
\and Stefano Rubele\inst{1,3}
\and Maria-Rosa Cioni\inst{4}
\\ (for the VMC Team)}
\institute{
 Osservatorio Astronomico di Padova -- INAF,
	Vicolo dell'Osservatorio 5, I-35122 Padova, Italy \and
 Universidade de S\~ao Paulo, IAG, Rua do Mat\~ao 1226, 
Cidade Universit\'aria, S\~ao Paulo 05508-900, Brazil \and
 Dipartimento di Astronomia, Universit\`a di Padova,
	Vicolo dell'Osservatorio 2, I-35122 Padova, Italy \and
Center for Astrophysics Research, University of Hertfordshire, 
Hatfield AL10 9AB, UK
}

\offprints{L. Kerber \\ e-mail: leandro.kerber@oapd.inaf.it}

\date{Received 9 October 2008 / Accepted}

%%%%%%%%%%%%%%%%%%%%%%%%%%%%%%%%%%%%%%%%%%%%%%%%%%%%%%%%%

\abstract
% Context
{The VISTA near infrared survey of the Magellanic System (VMC) will
provide deep $YJK_{\rm s}$ photometry reaching stars in the oldest
turn-off point all over the Magellanic Clouds (MCs).
%} 
% Aims
%{
As part of the preparation for the survey, we aim to access the
accuracy in the Star Formation History (SFH) that can be expected from
VMC data, in particular for the Large Magellanic Cloud (LMC).
%}
% Methods
%
To this aim, we first simulate VMC images containing not only the LMC
stellar populations but also the foreground Milky Way (MW) stars and
background galaxies. The simulations cover the whole range of density
of LMC field stars. We then perform aperture photometry over these
simulated images, access the expected levels of photometric errors and
incompleteness, and apply the classical technique of
SFH-recovery based on the reconstruction of colour-magnitude diagrams
(CMD) via the minimization of a chi-squared-like statistics. We verify
that the foreground MW stars are accurately recovered by the
minimization algorithms, whereas the background galaxies can be
largely eliminated from the CMD analysis due to their particular
colours and morphologies.
%}
% Results
%{
We then evaluate the expected errors in the recovered star formation
rate as a function of stellar age, SFR$(t)$, starting from models with
a known Age--Metallicity Relation (AMR). It turns out that, for a
given sky area, the random errors for ages older than $\sim0.4$ Gyr
seem to be independent of the crowding; this can be explained by a
counterbalancing effect between the loss of stars due to a decrease in
the completeness, and the gain of stars due to an increase in the
stellar density. For a spatial resolution of $\sim0.1\,\sqdeg$, the
random errors in SFR$(t)$ will be below 20\% for this wide range of
ages. On the other hand, due to the smaller stellar statistics for
stars younger than $\sim0.4$~Gyr, the outer LMC regions will require
larger areas to achieve the same level of accuracy in the SFR$(t)$. If
we consider the AMR as unknown, the SFH-recovery algorithm is able to
accurately recover the input AMR, at the price of an increase of
random errors in the SFR$(t)$ by a factor of about 2.5. Experiments of
SFH-recovery performed for varying distance modulus and reddening
indicate that these parameters can be determined with (relative)
accuracies of $\Delta\dmo\sim0.02$~mag and $\Delta\ebv\sim0.01$ mag,
for each individual field over the LMC. The propagation of these
latter errors in the SFR$(t)$ implies systematic errors below 30\%.
%}
%Conclusions
%{
This level of accuracy in the SFR$(t)$ can reveal important imprints
in the dynamical evolution of this unique and nearby stellar system,
as well as possible signatures of the past interaction between the MCs
and the MW.
}


\authorrunning{Kerber et al.}
\titlerunning{SFH Recovery of the LMC from VMC Survey}

\maketitle

%%%%%%%%%%%%%%%%%%%%%%%%%%%%%%%%%%%%%%%%%%%%%%%%%%%%%%%%%%%
\section{Introduction}
\label{sec_intro}

Determining the star formation histories (SFH) of the Magellanic
Clouds (MC) is one of the most obvious goals in the study of nearby
galaxies, for a series of reasons.  First, this SFH does likely keep
record of the past interactions between both Clouds and the Milky Way
\citep{Olsen99, Holtzman_etal99, Smecker-Hane_etal02, HZ04},
which are still to be properly unveiled \citep{KvdMA06a, KvdMA06b,
Besla_etal07, PPO08}. Detailed SFH studies may also provide unvaluable
hints on how star formation is triggered and proceeds in time, from
the smallest to galactic-size scales, and how these processes depend
on dynamical effects \citep[e.g.][]{HaZa07,Harris08}.

The Magellanic Clouds are also a rich laboratory for studies of
star formation and evolution, and the calibration of primary
standard candles, thanks to the simultaneous presence of a wide
variety of interesting objects such as red clump giants, Cepheids, RR
Lyrae, long period variables, carbon stars, planetary nebulae, the
tip of the red giant branch (RGB), dust-enshrouded giants, pre-main
sequence stars, etc. Although the system contains several hundreds of
star clusters for which age and metallicity can be measured, the bulk
of the interesting stellar objects are actually in the field and
irremediably mixed by the complex SFH, and also partially hidden by
the presence of variable and patchy extinction across the
MCs. Unveiling this complex SFH may help in calibrating stellar
properties -- like luminosities, lifetimes, periods, chemical types,
etc. -- as a function of age and metallicity.

In the last two decades, many authors demonstrated that recovering the
SFH of the MC from optical photometry is indeed feasible and well
worth of the effort. Such works are, usually, based either on deep
Hubble Space Telescope (HST) photometry reaching the oldest main
sequence turn-off for small MC areas \citep[e.g.][]{Gallagher_etal96,
Holtzman_etal99, Olsen99, Elson_etal97, Smecker-Hane_etal02,
Ardeberg_etal97, Dolphin_etal01, Javiel_etal05}, or on relatively
shallow ground-based photometry covering larger areas over the MCs
\citep{Stappers_etal97, GH92, HZ01, HZ04}. Only in very few cases
\citep[e.g.][]{Gallart_etal04, Noel_etal07} have the ground-based
optical photometry been deep enough to reach the oldest main sequence
turn-offs.

The VISTA Survey of the Magellanic System\footnote{See
http://www.vista.ac.uk and \\
http://www.star.herts.ac.uk/$\sim$mcioni/vmc/ \\ for further
information.} \citep[VMC, see][Cioni et al., in
preparation]{Cioni_etal07} is an ESO public survey project which will
provide, in the next 5 years, critical near-infrared data aimed --
among other goals -- to improve upon present-day SFH determinations.
This will hopefully open the way to a more complete understanding of
how star formation relates to the dynamical processes under way in the
system, and to a more accurate calibration of stellar models and
primary standard candles. Regarding the SFH, the key contributions of
the VMC survey will be: (1) It will provide photometry reaching as
deep as the oldest main sequence turn-off {\em over the bulk of} the
MC system, as opposed to the tiny regions sampled by HST, and the
limited area covered by most of the dedicated ground-based
observations. (2) VMC will use the near-infrared $YJK_{\rm s}$
passbands, hopefully reducing the errors in the SFH-recovery due to
variable extinction across the MCs.
%(3) The
%time-resolved photometry of variable stars, combined with
%previously-available data, will contribute to constrain the MC
%geometry, consequently limiting the parameter space to be explored by
%the SFH-recovery methods. Moreover, it is worth mentioning that VMC
%will have a much improved resolution as compared to previous
%near-infrared surveys (2MASS and DENIS), providing the only
%counterpart to existing optical surveys at a similar sensitivity.

On the other hand, the use of near-infrared instead of optical filters
will bring along some complicating factors, like an higher degree of
contamination of the MC photometry by foreground stars and background
galaxies, and the extremely high noise contributed by the sky,
especially in the $K_{\rm s}$ band.

Indeed, VMC will be the first near-infrared wide-area survey to
provide data suitable for the classical methods of
SFH-recovery\footnote{The previous attempts of \citet{Cioni_etal06a,
Cioni_etal06b} based on $IJK_{\rm s}$ data, were based on the shallow
observations from DENIS and 2MASS, which are limited to the upper RGB
and above.  Consequently, they could access the general trends in the
mean age and metallicity across the MCs, but not the detailed
age-resolved SFH.}. With the new space-based near-infrared cameras
(the HST/WFC3 IR channel, and JWST) and ground-based adaptive optics
facilities, observations similar to VMC ones will likely be available
for many nearby galaxies. VMC may become the precursor of detailed
SFH-recovery in the opening window of near-infrared
wavelengths. Demonstrating the feasibility of VMC goals, therefore, is
of more general interest.

Another particularity of the VMC survey is that, once started, its
data flow will be so huge that algorithms of analysis have better to
be prepared in advance, in the form of semi-automated
pipelines. Similar approaches have been/are being followed by some
ambitious nearly-all-sky (SDSS, 2MASS), micro-lensing (MACHO, OGLE,
EROS), and space astrometry (e.g. Hipparcos, GAIA) surveys.

In this paper, we describe part of the preparatory work for the
derivation of the SFH from VMC data, which can be summarised in the
following way: First we simulate the VMC images for the LMC
(Sect.~\ref{sec_simul}), where we later perform the photometry and
artificial stars tests (Sect.~\ref{sec_photom}) that allow us to
access the expected levels of photometric errors, completeness, and
crowding, and the contamination by foreground stars and background
galaxies. We then proceed with many experiments of SFH-recovery
(Sect.~\ref{sec_sfh}), evaluating the uncertainties in the derivation
of the SFH as a function of basic quantities such as the stellar
density over the LMC, the area included in the analysis, and the
adopted values for the distance and reddening. Doing this, we are able
to present the expected random and systematic errors in the
space-resolved SFH.  Such information may be useful to plan
complementary observations and surveys of the LMC in the next few
years. Further papers will present the perspectives for
the Small Magellanic Cloud (SMC), %for which \citet{HZ04} determined a
%space-resolved SFH using $UBVI$ data, 
as well as better explore the
effect in the recovered SFH due to the uncertainties associated with
the MC geometry, differential reddening, initial mass function,
fraction of binaries, etc.

%%%%%%%%%%%%%%%%%%%%%%%%%%%%%%%%%%%%%%%%%%%%%%%%%%%%%%%%%%%
\section{Simulating VMC data}
\label{sec_simul}

Our initial goal is to obtain realistic simulations of VMC images,
containing all of the objects that are known to be present towards the
MCs and likely to be detectable within the survey depth limits. These
objects are essentially stars belonging to the MW and the MCs, and
background galaxies. Moreover, an essential component of the images is
the high signal from the infrared sky. Each one of these components
will be detailed below. Diffuse objects such as emission nebulae and
star clusters will, for the moment, be ignored.

\subsection{VISTA and VMC specifications}
\label{vista_and_vmc}

%\begin{table}%[!ht]
%%\centering
%\caption{VMC mean and end-of-survey specifications}
%\label{tab:filtcon}
%\begin{tabular}{l|llll}
%\hline
%Parameter & $Y$ & $J$ & $K_{\rm s}$ & units\\
%\hline
%sky brightness &   17.2    &  16.0    &  13.0    &  mag~arcsec$^{-2}$\\
%%$S_{\rm sky}$  &   190.6   &  762.1   &  5784.6  &  e$^{-}$s$^{-1}$pix$^{-1}$\\
%$S_{\rm sky}$  &   1658.5  &  6631.6  &  50335.9 &  e$^{-}$s$^{-1}$arcsec$^{-2}$\\
%$t_{\rm exp}$   &   3000    &  2400    &  10800   &  s\\
%$z_{\rm mag}$ &   25.249  &  25.554  &  24.755  &  mag\\
%seeing	& 1.0	& 0.9	& 0.8	& arcsec\\
%seeing (center tiles)	& 0.8	& 0.7	& 0.6	& arcsec\\
%mag at SNR=10 	& 21.9	& 21.4	& 20.3	& mag \\
%%$z_{\rm mag,2}$ &   33.942  &  34.005  &  34.839  &  mag\\
%%$S_{\rm sky,\,tot}$  &   5.718e5 &  1.829e6 &  6.247e7 &  e$^{-}$pix$^{-1}$\\
%\hline
%\end{tabular}
%%where $z_{\rm mag,2} = z_{\rm mag,1} + 2.5\,\log t_{\rm exp}$,
%%$S_{\rm sky,\,tot} = S_{\rm sky}\,t_{\rm exp}$ (total signal 
%%from the sky in the final images), and
%%$m=-2.5\log(S/t_{\rm exp}) + z_{\rm mag,1}$ or $m = -2.5\log S + z_{\rm mag,2}$.\\
%\end{table}

VMC will be performed with the VIRCAM camera mounted at the 4m VISTA
telescope at ESO's Paranal Observatory in Chile.
VIRCAM has 16 2048$\times$2048 detectors which, with
the image scale of 0.339~$\arcsec$ per pixel on average, cover a sky
area of 0.037~$\sqdeg$ each. The basic mode of the observations will
be to perform 6 exposures (paw-prints) with the subsequent construction
of $1.0\times1.5$~$\sqdeg$ tiles. In the following, we will adopt the
area of each detector (i.e. 0.037~$\sqdeg$) as the basic unit of our
simulations.

The specifications of the VMC survey will be detailed in another paper 
(Cioni et al., in preparation).
%Table~\ref{tab:filtcon} presents some specifications of VMC and its
%observing site. 
For our aims, suffice it to mention the following:
Despite for the crowded fields, it is expected that
the observations will be sky-noise dominated. The mean sky brightness
at Cerro Paranal is of 17.2, 16.0, 13.0 mag~arcsec$^{-2}$ in $YJK_{\rm
s}$, respectively. The required seeing is of 1.0~$\arcsec$ (FWHM) in the 
$Y$ band, being the most crowded
regions observed in nights with seeing better than 0.8~$\arcsec$.
%For simplicity, the above-mentioned mean values have been adopted in all 
%of our simulations. 
%The total exposure times $t_{\rm exp}$ for each tile will be of
%3000, 2400, 10800 seconds in $YJK_{\rm s}$, 
%necessary to produce a
The targetted signal-to-noise ratio (SNR) is equal to 10 at magnitudes
of 21.9, 21.4, 20.3 mag respectively. The photometric zero-points in
our simulations were fixed via the VISTA exposure time calculator, so
as to be consistent with these values.
Considering these survey limits, in our simulations we
include all objects brighter than $K_{\rm s}=22.5$, which at the LMC
distance correspond to a stellar mass of $\sim 0.8$~\Msun\ in the main
sequence turn-off.

\begin{figure}
\resizebox{\hsize}{!}{\includegraphics{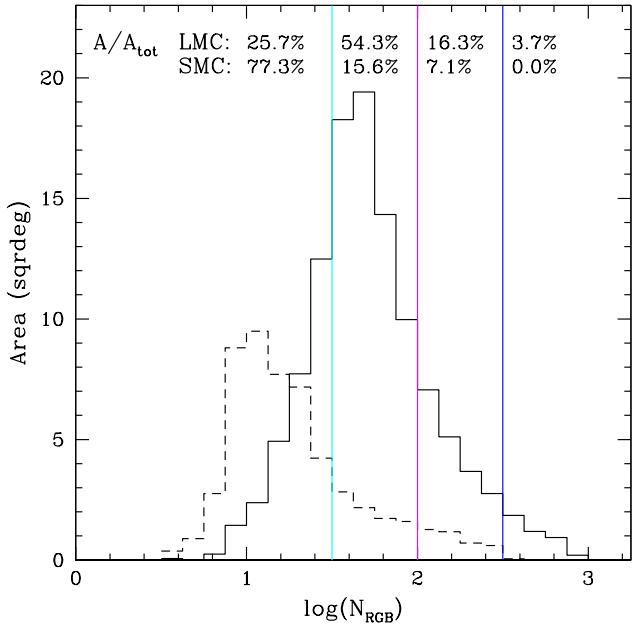}}
\caption{
The area VMC will likely cover in the LMC (solid line) and SMC (dashed line) 
as a function of the surface density of RGB stars, $N_{\rm RGB}$. 
The fraction of the covered area in each MC,  
for four ranges of density, is also shown in the top of the figure.}
%, where the colours of the horizontal lines are in
%accordance to the density levels shown in Fig.~\ref{vmc_fig} .  }
\label{vmc_distdens}
\end{figure}

VMC tiles will cover most of the Magellanic System,
%Figure \ref{vmc_fig} shows the likely distribution of the 110 VMC
%tiles across the Magellanic System. They cover 110 $\sqdeg$ in the
%LMC, 45 $\sqdeg$ in the SMC, 20 $\sqdeg$ in the Bridge and 3 $\sqdeg$
%in Stream, 
summing to a total area of 184~$\sqdeg$ \citep[see][ Cioni et al., in
preparation, for details]{Cioni_etal07}. Fig.~\ref{vmc_distdens} shows
a histogram of the total area to be observed as a function of the
density of upper RGB stars, $N_{\rm RGB}$, which is defined as the
number of 2MASS stars inside a box in the $K_{\rm s}$ vs. $J-K_{\rm
s}$ CMD ($ 0.60 \leq J-K_{\rm s} \leq 1.20$ and $12.00
\leq K_{\rm s} \leq 14.00$ for the LMC and $12.30 \leq K_{\rm s} \leq
14.30$ for the SMC), for each unit area of 0.05 $\sqdeg$. Notice the
higher mean and maximum stellar densities of the LMC, as compared to
the SMC. The stellar densities vary over an interval of about 2.5~dex.

%Fig.~\ref{vmc_fig}
%well illustrates the wide range of stellar densities -- varying over
%an interval of about 2.5~dex -- that will be covered by
%VMC.  The points in
%this figure (top panel) are 2MASS stars selected in a box inside the
%RGB ($ 0.60 \leq J-K_{\rm s} \leq 1.20$, and $12.50 \leq K_{\rm s}
%\leq 12.75$ for the LMC and $12.80 \leq K_{\rm s} \leq 13.05$ for the
%SMC).  For further use in this paper, we define the quantity $N_{\rm
%RGB}$, which is the surface density of stars observed in the RGB
%within $\sim 2$ magnitudes of its tip.  In practice $N_{\rm RGB}$ is
%obtained counting 2MASS stars inside a box in the $K_{\rm s}$
%vs. $J-K_{\rm s}$ CMD ($ 0.60 \leq J-K_{\rm s} \leq 1.20$ and $12.00
%\leq K_{\rm s} \leq 14.00$ for the LMC and $12.30 \leq K_{\rm s} \leq
%14.30$ for the SMC) for an area of 0.05 $\sqdeg$.  Fig.~\ref{vmc_fig}
%well illustrates the wide range of stellar densities -- varying over
%an interval of about 2.5~dex -- that will be covered by
%VMC. Fig.~\ref{vmc_distdens} shows a histogram of the total area to be
%observed as a function of $N_{\rm RGB}$, delimiting the regions in
%correspondence with the density levels of Fig.~\ref{vmc_fig}. Notice
%the higher mean and maximum stellar densities of the LMC, as compared
%to the SMC.
%
%\begin{figure}
%\resizebox{\hsize}{!}{\includegraphics{vmcsky_dens_cut.ps}}
%\caption{Distribution of VISTA tiles across the Magellanic System.
%The density of 2MASS upper RGB stars ($N_{\rm RGB}$) in a logarithmic
%scale is represented by the colour scale (see text and
%Fig.\ref{vmc_distdens} for details).}
%\label{vmc_fig}
%\end{figure}

\subsection{Stars in the UKIDSS photometric system}

Since VISTA is still being commissioned at the time of this writing, the
throughputs of VISTA filters, camera, and telescope are still not
available. It is however clear that the VISTA photometric system will
be very similar to the UKIDSS one, with the differences being mainly
in the higher performance of VISTA, and in the fact that VISTA will
use a $K$-short filter ($K_{\rm s}$) similar to the 2MASS one.
%This is
%illustrated in Fig.~\ref{fig_filters}, which compares a very
%preliminar set of filter transmission curves for VISTA (kindly
%provided by Jim Emerson) with UKIDSS (Hewett et al. 2005) and 2MASS
%(Cutri et al. ????) ones. Additionally, the figure includes the SDSS
%$z$-band (York?????), in order to illustrate the particular position
%of the $Y$-band, which has no counterpart in wide-field public surveys
%previous to UKIDSS.
%
%\begin{figure}
%\resizebox{\hsize}{!}{\includegraphics{filtri_vista.ps}}
%\caption{The preliminary filter throughputs for VISTA, as compared to
%UKIDSS and 2MASS + SDSS $z$ filters (upper panels), and to the
%intrinsic spectrum emitted by an M5 giant, the Sun, and Vega (bottom
%panel). All curves have been renormalised at its maximum of
%transmission or flux. CAN WE PUBLISH THIS FIGURE? THIS HAS TO BE ASKED 
%TO JIM EMERSON PRIOR TO SUBMISSION. }
%\label{vmc_filters}
%\end{figure}

Given the present situation, we have so far used the UKIDSS system as
a surrogate of the future VISTA one. Tests using the preliminary VISTA
filter curves (Jim Emerson, private communication) indicate very small
differences in the synthetic photometry, typically smaller than
0.02~mag, between VISTA and UKIDSS. \footnote{Throughout this
paper, we will name the 2.2~$\mu$m filter as $K_{\rm s}$ when
referring to VISTA, 2MASS and DENIS, and as $K$ when referring to our
simulations and to UKIDSS data. Notice however that, for all practical
purposes, the actual difference between these filters is not a matter
of concern.}

\begin{figure}
\resizebox{\hsize}{!}{\includegraphics{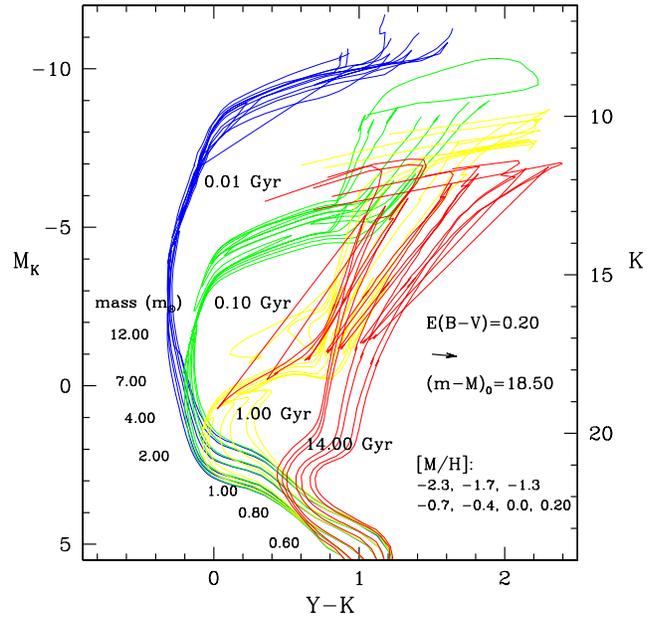}}
\caption{A series of \citet{Marigo_etal08} isochrones in the UKIDSS 
photometric system.  The figure shows the absolute ($M_{K}$, $\yk$)
CMD as well as the apparent ($K$, $\yk$) one for a typical distance to
the LMC.  Stellar masses and isochrone ages and metallicities are also
indicated in the figure.  }
\label{fig_isoc}
\end{figure}

Stellar isochrones in the UKIDSS system have been recently provided by
\citet{Marigo_etal08}\footnote{http://stev.oapd.inaf.it/dustyAGB07}. The
conversion to the UKIDSS system takes into account not only the
photospheric emission from stars, but also the reprocessing of their
radiation by dusty shells in mass-losing stars, as described in
\citet{Marigo_etal08}. The filter transmission curves and zero-point
definitions come from \citet{Hewett_etal06}. The stellar models in use
are composed of \citet{Girardi_etal00} tracks for low- and
intermediate-mass stars, replacing the thermally pulsing asymptotic
giant branch (AGB) evolution with the \citet{MG07} ones. In this
paper, these models are further complemented with white and brown
dwarfs as described in \citet[][{also Zabot et al., in
preparation}]{Girardi_etal05}, and with the \citet{Bertelli_etal94}
isochrones for masses higher than 7~$M_\odot$. Fig.~\ref{fig_isoc}
presents some of the \citet{Marigo_etal08} isochrones in the $M_K$
vs. $\yk$ diagram, for a wide range in age and metallicity. As can be
readily noticed, the isochrones contain the vast majority of the
single objects that can be prominent in the near-infrared observations
of the LMC, going from the lower Main-Sequence (MS) stars, 
up to the brightest AGB stars and red supergiants. 
The stellar masses in the MS and the
apparent magnitude for a typical LMC distance, $\dmo = 18.50$
\citep{Clementini_etal03, Alves04, Schaefer08} are also indicated in
this figure. We recall that our models contain, in addition, the
very-low mass stars, brown dwarfs and white dwarfs, which are
important in the description of the foreground MW population
\citep{Marigo_etal03}.

The interstellar extinction coefficients adopted in this work do also
follow from \citet{Marigo_etal08}: $A_Y=0.385\,A_V$, $A_J=0.283\,A_V$,
and $A_K=0.114\,A_V$, which imply $\eyj=0.351\,\ebv$ and
$\eyk=0.931\,\ebv$.  They have been derived from synthetic photometry
applied to a G2V star extincted with \citet{Cardelli_etal89}
extinction curve. Although the approach is not the most accurate one
\citep[see][]{Girardi_etal08}, it is appropriate to the conditions of
moderate reddening ($\ebv \la0.2$~mag) typical of the Magellanic
Clouds.

The simulation of the input photometric catalogues and the generation
of artificial images are described in the next subsections. In brief,
the input catalogues for the LMC (Sect.~\ref{lmc_stars}) and the
foreground MW stars (Sect.~\ref{mw_foreground}) come from the
predictions made with the TRILEGAL code \citep{Girardi_etal05},
that simulates the photometry of resolved stellar populations following
a given distribution of initial masses, ages, metallicities, reddenings
and distances. The catalogues of background galaxies 
(Sect.~\ref{sec_galaxies}) are randomly drawn from UKIDSS
\citep{Lawrence_etal07}.  The simulation of images is performed with
the DAOPHOT and ARTDATA packages in IRAF\footnote{IRAF is distributed
by the National Optical Astronomy Observatory, which is operated by
the Association of Universities for Research in Astronomy (AURA) under
cooperative agreement with the National Science Foundation.} 
(Sect.~\ref{simulating_galaxies}), always respecting the photometric
calibration and expected image quality required by the VMC survey.

\subsection{The LMC stars}
\label{lmc_stars}

The stellar populations for the LMC are simulated as an ``additional
object'' inside the TRILEGAL code \citep{Girardi_etal05}, where the
input parameters for a field are:
\begin{itemize}
\item the star formation rate as a function of stellar age, SFR$(t)$;
\item the stellar AMR, $Z(t)$ or [M/H]$(t)$;
\item the total stellar mass, $M_{\rm{tot,LMC}}$;
\item the distance modulus, $\dmo$;
\item the reddening, $\ebv=3.1\,A_V$;
\item the Initial Mass Function (IMF), $\psi(M_{\rm i})$;
\item the fraction of detached unresolved binaries, $f_{\rm bin}$.
%\item the equatorial coordinates (RA, DEC). 
%\noindent
%- Area field (A$_{\rm field}$). 
\end{itemize}

As commented before, for convenience we are simulating an area of
0.037~$\sqdeg$, equivalent to a $2048\times2048$ VIRCAM detector.  The
value of $M_{\rm{tot,LMC}}$ is suitably chosen such as to generate the
total number of RGB stars observed by 2MASS, $N_{\rm RGB}$, inside
this same area.

In the LMC simulations presented in this paper, we adopt an input AMR
consistent with the one given by stellar clusters
\citep{Olszewski_etal91, MG03, Grocholski_etal06, Kerber_etal07} and
field stars \citep{Cole_etal05, Carrera_etal08}, together with a constant
SFR$(t)$. Since the SFR$(t)$ in the LMC is clearly spatial dependent
\citep{Holtzman_etal99, Smecker-Hane_etal02, Javiel_etal05}, the assumption of
a constant SFR$(t)$ shall be considered as just a way to ensure a uniform
treatment for all stellar populations over the LMC.
%The case of non-constant SFR$(t)$ will be analysed later 
%in this paper (?????).

In terms of distance we are initially using the canonical value of
$\dmo=18.50$ \citep{Clementini_etal03, Alves04, Schaefer08} also
adopted by the HST Key Project to measure the Hubble constant 
\citep{Freedman_etal01}, whereas for the reddening we are assuming a
value of $\ebv=0.07$, typical for the extinction maps from
\citet{Schlegel_etal98}. For real VMC
images, these two parameters are expected to be free parameters, since
the LMC presents disk-like geometries with a significant inclination
($\sim$ 30--40 deg) \citep{vdMC01, vanderMarel_etal02, Nikolaev_etal04} 
and non-uniform extinction \citep{Zaritsky_etal04, Subramaniam05, IB07}.

Finally the assumed values for the remaining inputs are: the 
\citet{Chabrier01} lognormal IMF
\footnote{With a slope $\alpha \sim -2.3$ for $0.8 < m/\Msun < 5.0$
and $\alpha \sim -3.0$ for $m > 5.0 \Msun$, where the Salpeter slope
is $\alpha=-2.35$.}
, and $f_{\rm bin}=30$\,\% with a constant mass
ratio distribution for $m_{2}/m_{1} > 0.7\:$\footnote{This is the mass
ratio interval in which the secondary significantly affects the
photometry of the system.}. There are no strong reasons to expect
significant deviations for these choices, especially for the IMF since
we are dealing with stars with masses approximately between 0.8 and
12.0 $\Msun$ where the IMF slope seems to be universal 
and similar to the Salpeter one \citep{Kroupa01,Kroupa02}.
Concerning the fraction of binaries, our choice is
consistent with the values found for the stellar clusters in the LMC
\citep{Elson_etal98, Hu_etal08}. For the moment, these will be
considered as fixed inputs. Further papers will use simulations in
order to quantify the systematic errors in the recovered SFH
introduced by the uncertainties related to these choices.

\subsection{The Milky Way foreground}
\label{mw_foreground}

The MW foreground stars are simulated using the TRILEGAL code as
described in \citet{Girardi_etal05}. Towards the MCs, the simulated
stars are located both in a disk which scale-height increasing with
age, and in a oblate halo component. Diffuse interstellar reddening
within 100~pc of the Galactic Plane is also considered, although it
affects little the near-infrared photometry.

In \citet{Girardi_etal05}, it has been shown that for off-plane
line-of-sights, TRILEGAL predicts star counts accurate to within about
15\% over a wide range of magnitudes and down to $J\simeq20.5$ and
$K\simeq18.5$. This accuracy is confirmed by the $K\la20.5$
observations of \citet{Gullieuszik_etal08} for a field next to the
Leo~II dwarf spheroidal galaxy. Moreover, \citet{Marigo_etal03} shows
that TRILEGAL describes very well the position of the three ``vertical
fingers'' observed in 2MASS $K$ vs. $J-K_{\rm s}$ diagrams.
Similarly-comforting comparisons with UKIDSS data (including the $Y$
band) are presented in Sect.~\ref{sec_comp_ukidss} below.

Although predicting star counts with an accuracy of about 15\% may be
good enough for our initial purposes, we are working to further
improve this accuracy: In short, we are applying the minimisation
algorithm described in \citet{Vanhollebeke_etal08} -- which was
successfully applied to the derivation of Bulge parameters using data
for inner MW regions -- to recalibrate the TRILEGAL disk and halo
models. It is likely that before VMC starts, foreground star counts
will be predicted with accuracies of the order of 5\%.

\subsection{The background galaxies}
\label{sec_galaxies}

In order to simulate the population of galaxies background to the MCs,
we make use of the large catalogues of real galaxies obtained by the
UKIDSS Ultra-Deep \citep[UDS;][]{UKIDSSuds} and Large Area Surveys
\citep[LAS;][]{UKIDSSlas}, from their Data Release 3 (December 2007). The
LAS includes data for an area of 4000~$\sqdeg$ down to $K=18.4$, for
$YJK$ filters, whereas the UDS includes an area of 0.77~$\sqdeg$
observed down to $K\sim23$, but only for $JHK$ passbands.

In our input catalogue for each simulation, we include the number of
UKIDSS galaxies expected for our total simulated area. More precisely,
we randomly pick up from the UDS catalogue, a fraction of galaxies
given by the ratio between the areas covered by UDS and by our image
simulation.  From the catalogue, we extract their $J$ and $K$
magnitudes, and morphological parameters (position angle, size, and
axial ratio). In this way, our simulations respect the observed
$K$-band luminosity function of galaxies, and their $J-K$ colour
distribution, down to faint magnitudes. The $Y$-band magnitudes,
instead, have been assigned in the following way: we take the $J-K$
colour of each galaxy in the UDS, and then randomly select a galaxy
from the LAS which has the most similar $J-K$ (within 0.2 mag), and
take its $Y-J$. This means that the $Y-J$ vs. $J-K$ relation from LAS
is being extrapolated down to deeper magnitudes\footnote{This is of
course a crude approximation since deeper surveys sample larger galaxy
redshifts. However, it is justified by the lack of deep-enough $Y$
data, and by the little impact that such faint galaxies have in our
stellar photometry (see Sect.~\ref{sec_comp_ukidss}).}.
%Galaxies are added at random positions in the image, with the
%\textbf{mkobject} IRAF task, and adopting two simple typical profiles:
%exponential disks for spirals and de Vaucouleurs $r^{1/4}$ laws for
%ellipticals. 

\subsection{Simulating images}
\label{simulating_galaxies}

Once defined the input catalogues for stars and galaxies we simulated 
the images inside IRAF, in accordance with the VISTA and VMC specifications
(see Sect.\ref{vista_and_vmc}).
The basic sequence of steps (and the IRAF {\it task}) performed for 
a given filter is the following:
\begin{enumerate}
\item
Definition of the image size ({\it rtextimage}) and introduction 
of the sky brightness and noise ({\it mknoise});
\item
simulation of a Gaussian stellar profile ({\it gauss}) 
respecting the expected seeing for an image of a photometric 
calibrated (using the VISTA ETC v1.3) delta function with a 
known number of electrons;
\item
addition of the LMC and MW stars in the sky images ({\it addstars}) 
following the previous calibrated Gaussian stellar profile
with random poissonian errors in the number of electrons; 
\item
addition of galaxies ({\it mkobject}) in the previous image  
respecting all information concerning the morphological type, position angle, 
size and axial ratio. 
\end{enumerate}

\begin{figure}
\resizebox{\hsize}{!}{\includegraphics{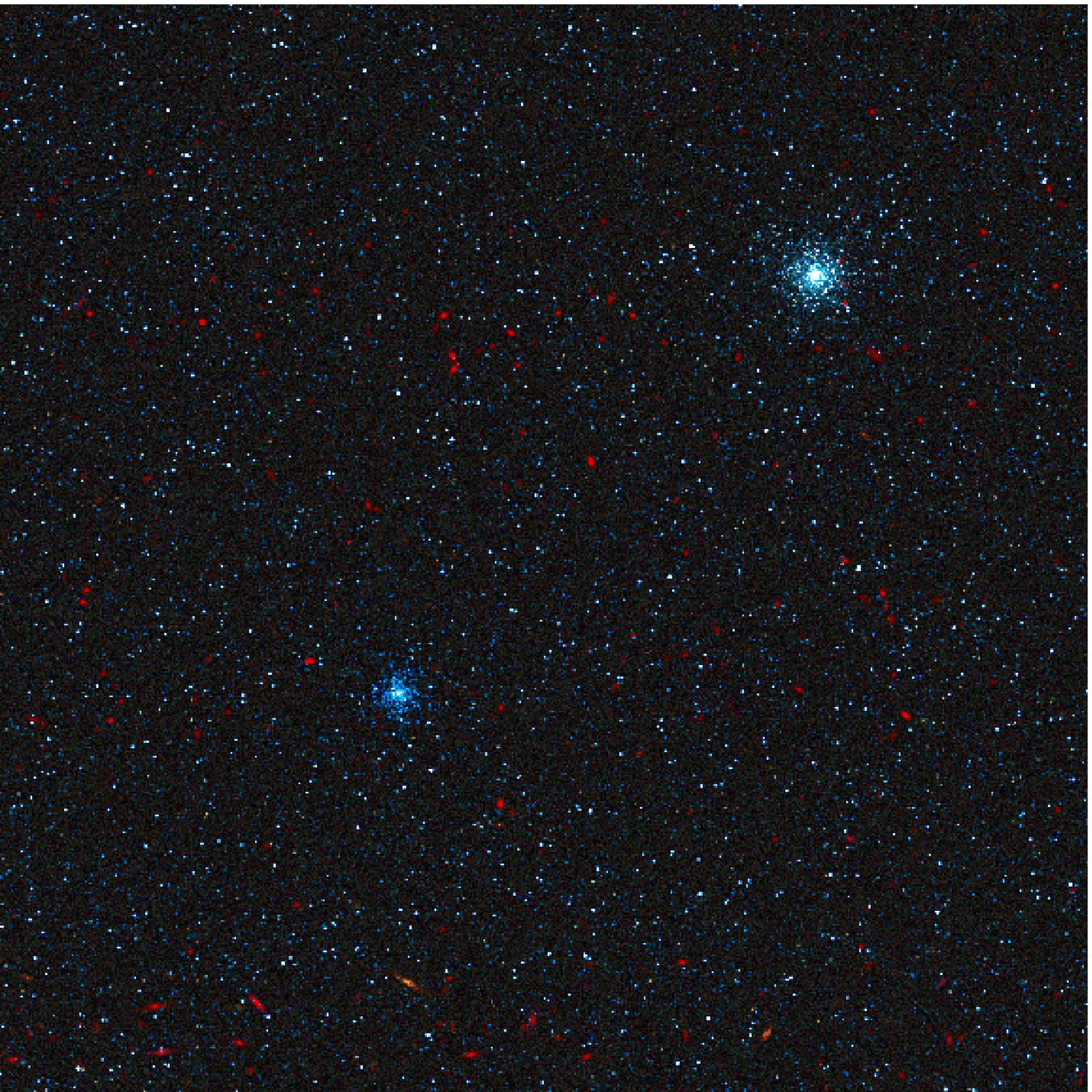}}
\smallskip
\begin{minipage}{0.495\columnwidth}
\resizebox{\hsize}{!}{\includegraphics{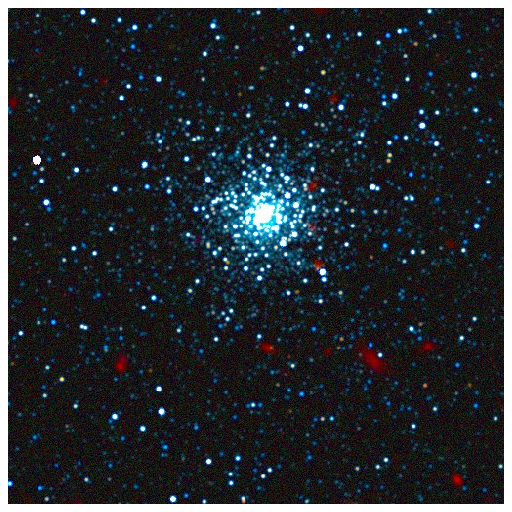}}
\end{minipage}
\hfill
\begin{minipage}{0.495\columnwidth}
\resizebox{\hsize}{!}{\includegraphics{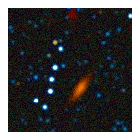}}
\end{minipage}
\caption{An image simulation for the area next to the star cluster
NGC1805 ($\alpha=5.03$~h, $\delta=-66.07^\circ$), for a single
$2048\times2048$ array detector of VIRCAM. This is a false-color image
where blue-green-red colours were associated to the $YJK$ filters,
respectively. The location corresponds to a $\log N_{\rm RGB} \sim
2.00$ in Fig.~\ref{vmc_distdens}. The detector area corresponds to
0.0372~$\sqdeg$ ($11.6\times11.6$ arcmin) in the sky, which is about
1/40 of a single VIRCAM tile, and 1/5000 of the total VMC survey area.
The two small panels at the bottom present details of the simulated
stars, stellar clusters and galaxies for $2.9\times2.9$ arcmin and
$0.7\times0.7$ arcmin areas.  At the LMC distance the top panel
corresponds approximately to a box of $175\times175$ pc, whereas the
bottom panels correspond to $44\times44$ pc and $11\times11$ pc,
respectively (1 $\arcsec \sim 0.25$ pc).  }
\label{fig_image}
\end{figure}

To assure a uniform distribution of the objects in the image, stars
and galaxies are always added at random positions.
Fig.~\ref{fig_image} shows an example of image simulation for a
typical field in the LMC disk. The false-colour plot evidences the
colour and morphologic differences between stars and background
galaxies -- with the latter being significantly redder than the
former. In the same image, we have inserted two populous 
stellar clusters typical for the LMC with different ages, 
masses and concentration of stars (following a King's profile), 
just to illustrate our capacity to simulate also this kind of 
stellar object.

%%%%%%%%%%%%%%%%%%%%%%%%%%%%%%%%%%%%%%%%%%%%%%%%%%%%%%%%%%%

\section{Performing photometry on simulated data}
\label{sec_photom}

\subsection{Aperture and PSF photometry}

\begin{figure*}
\begin{minipage}{0.49\textwidth}
\resizebox{\hsize}{!}{\includegraphics{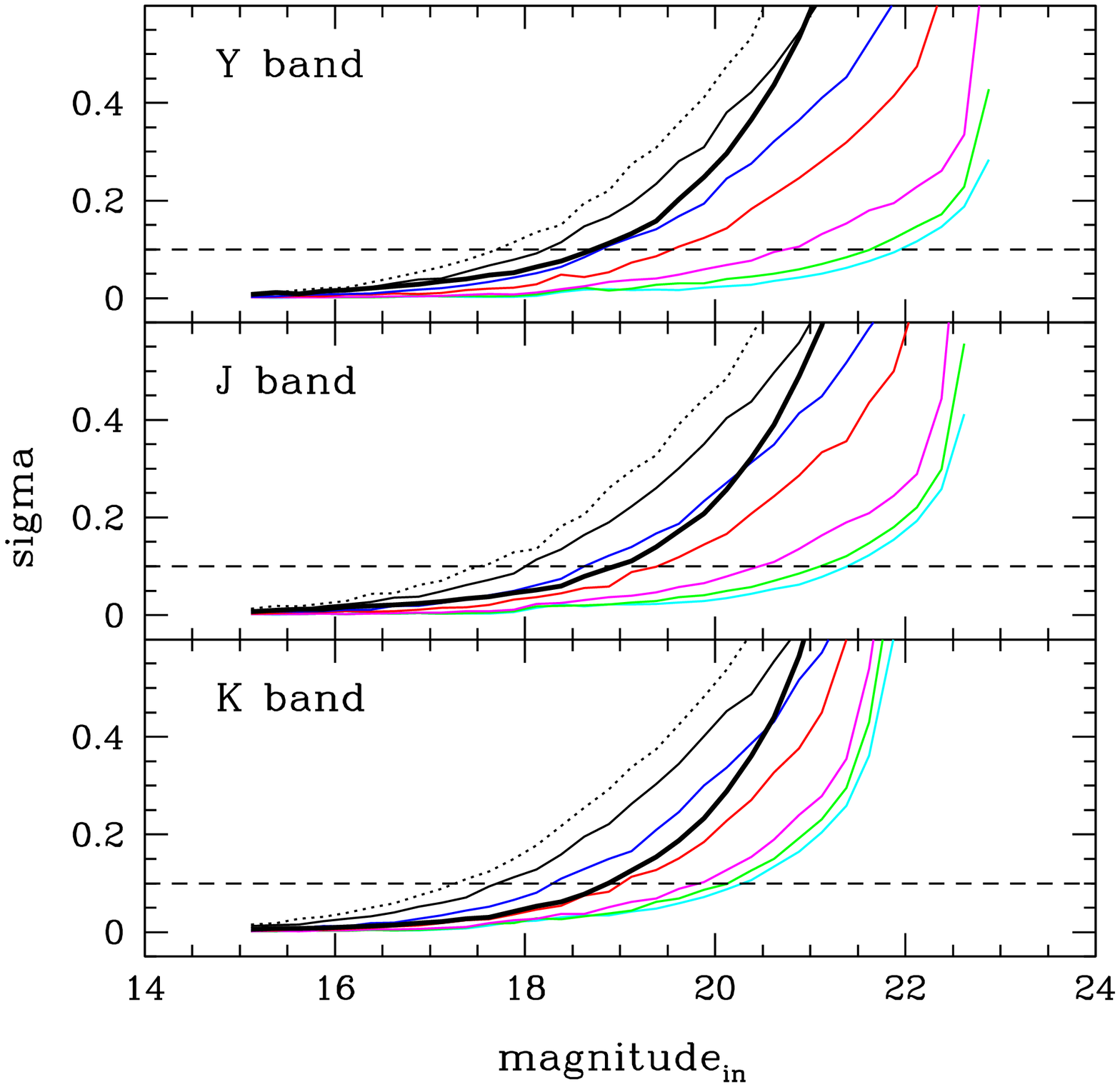}}
\end{minipage}
\hfill
\begin{minipage}{0.49\textwidth}
\resizebox{\hsize}{!}{\includegraphics{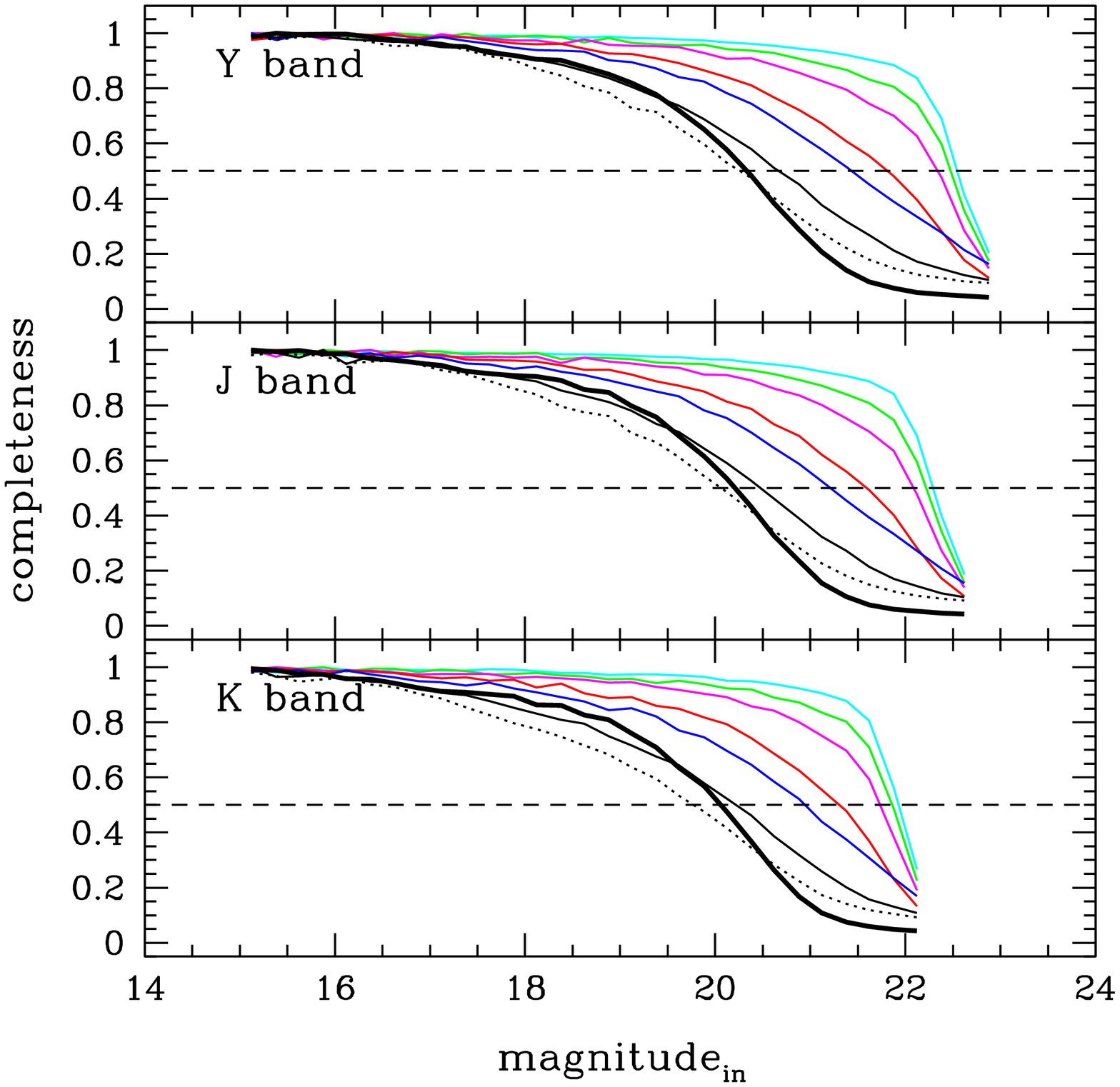}}
\end{minipage}
\hfill
\caption{
Photometric errors (left panels) and 
completeness curves (right panels) in the artificial 
$YJK$ images for the LMC for
different levels of crowding. The thin curves present the results for
the aperture photometry covering the entire expected range of density
of field stars ($\log N_{\rm RGB}=$ 1.50, 1.75, 2.00, 2.25, 2.50,
2.75, 2.90, see also Fig.~\ref{vmc_distdens}).  The
three highest density levels were simulated with the smallest values
for seeing required for the LMC centre.
The thick black line illustrates the results of performing PSF
photometry for the highest density level ($\log N_{\rm RGB}=2.90$).
The expected error in magnitude for a SNR=10 is shown by the dashed
line.}
\label{photo_all}
\end{figure*}

The IRAF DAOPHOT package was used to detect and to perform aperture
photometry in our simulated images. Candidate stars were detected
using {\it daofind}, with a peak intensity threshold for detection set
to $4\,\sigma_{\rm sky}$, where $\sigma_{\rm sky}$ corresponds to the
rms fluctuation in the sky counts. The aperture photometry was carried
out running the task {\it phot} for an aperture radius of 3~pix ($\sim
1.0 \arcsec$).

The photometric errors and completeness curves that come from this
aperture photometry in our simulated LMC images can be seen in
Fig.~\ref{photo_all}. 
The photometric errors in this case are estimated using the differences
between the input and output magnitudes; more specifically, for each
small magnitude bin we compute the half-width of the error 
distribution, with respect to the median,
that comprises 70\% of the recovered stars. The completeness instead
is simply defined as the ratio between total number of input stars,
and those recovered by the photometry package.
Fig.~\ref{photo_all} presents the
results for different simulations covering a wide range of density of
field stars in the LMC, from the outer disk regions to the centre
regions in the bar (see Fig.\ref{vmc_distdens}).
Notice that in these simulations we are following the requirement
that the most central and crowded regions ($\log N_{\rm RGB} \ge
2.50$) will be observed under excellent seeing conditions only.

It can be noticed that the VMC expected magnitudes at SNR=10 for
isolated stars ($Y=21.9$, $J=21.4$, $K=20.3$) is well recovered in the
simulations for the lowest density regions, attesting the correct
photometric calibration of our simulated images. For these regions the
50\% completeness level is reached at $Y\sim 22.5$, $J\sim
22.2$ and $K\sim 21.8$.

Crowding significantly affects in the quality of the aperture
photometry, making the stars measured in central LMC regions to appear
significantly brighter, and with larger photometric errors, than in
the outermost LMC regions. 
As shown in Fig.~\ref{photo_all}, crowding clearly starts to 
dominate the noise for fields with $\log N_{\rm RGB} \ga 2.00$ 
that correspond to about 20\% (7\%) of the total area covered in the LMC (SMC) 
(see Fig.~\ref{vmc_distdens}).
Therefore, PSF photometry is expected to be
performed whenever crowding will prevent a good aperture photometry
over VMC images.  The significant improvements that can be reach by a
PSF photometry are also illustrated by the thick black lines in
Fig.~\ref{photo_all}. These results
correspond to a PSF photometry applied to the LMC centre ($\log N_{\rm
RGB}=2.90$), where the PSF fitting and the photometry were done using
the IRAF tasks {\it psf} and {\it allstar}.

\begin{figure}
\resizebox{\hsize}{!}{\includegraphics{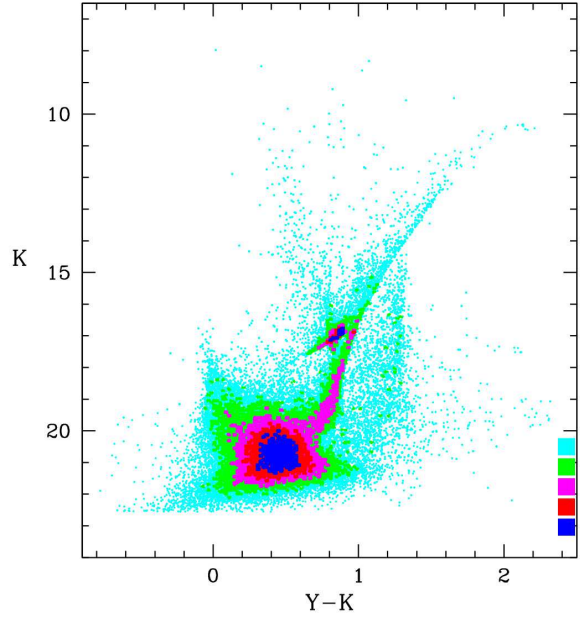}}
\caption{Example of ($K$, $\yk$) CMD from aperture photometry in a  
simulated image for the VMC survey.  The choices in the parameters
represent a field of $\sim0.1$~deg$^2$ with $\sim 10^{5}$ stars
($\log(N_{\rm RGB})=2.00$) following a constant SFR$(t)$ and a AMR
typical for the LMC clusters (see details in the text).  The colours
represent the density of points in a logarithmic scale.  The
information about approximated stellar masses, ages and metallicities
can be obtained from Fig.~\ref{fig_isoc}.  }
\label{cmd_hess}
\end{figure}

Figure \ref{cmd_hess} shows an example of CMD obtained from the
aperture photometry in a simulated field with an intermediate level of
density in the LMC. This figure reveals the expected CMD features --
and the wealth of information -- that will become available thanks to
the VMC survey: well evident are the AGB, red supergiants, RGB, red
clump (RC), Sub-Giant Branch (SGB) as well as the MS, from the
brightest and youngest stars down to the oldest turn-off point. In
comparison, the present-day near-infrared surveys of the MCs are
complete only for the most evolved stars -- excluding those in the
most crowded regions, and those highly extincted by circumstellar
dust. DENIS and 2MASS, for instance, are limited to $K_{\rm s}\la14$,
revealing the red supergiants, AGB and upper RGB, and including just a
tiny fraction of the upper MS \citep{NW00, Cioni_etal07}. IRSF
\citep{Kato_etal07} extends this range down to $K_{\rm s}\la16.6$,
which is deep enough to sample the RC and RGB bump, but not the SGB
and the low-mass MS.

\subsection{Comparison with UKIDSS data}
\label{sec_comp_ukidss}

Since the present work depends on simulations, it is important to
check if they reproduce the basic characteristics of real data already
obtained under similar conditions. UKIDSS represents the most similar
data to VMC to be available for the moment. Therefore, in the
following we will compare a simulated UKIDSS field with the real one.

For this exercise, we take the 0.21 $\sqdeg$ field towards Galactic
coordinates $\ell=-220\degr, b=40\degr$, which due to its similar distance from
the Galactic Plane as the MCs, offers a good opportunity to test the
expected levels of Milky Way foreground, and the galaxy background.

%%%%%%%%Figure %%%%
\begin{figure*}
\begin{minipage}{0.49\textwidth}
\resizebox{\hsize}{!}{\includegraphics{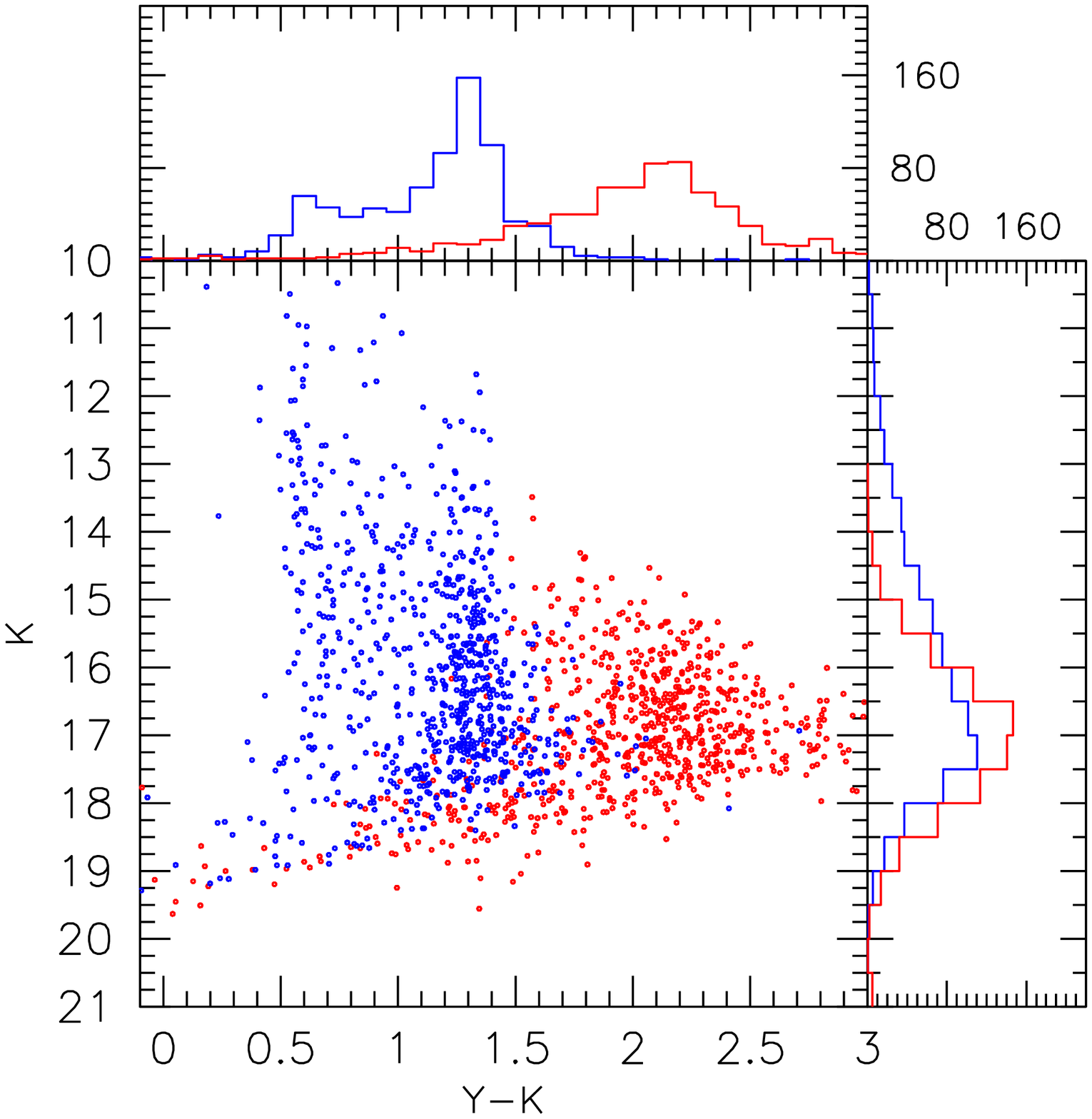}}
\end{minipage}
\hfill
\begin{minipage}{0.49\textwidth}
\resizebox{\hsize}{!}{\includegraphics{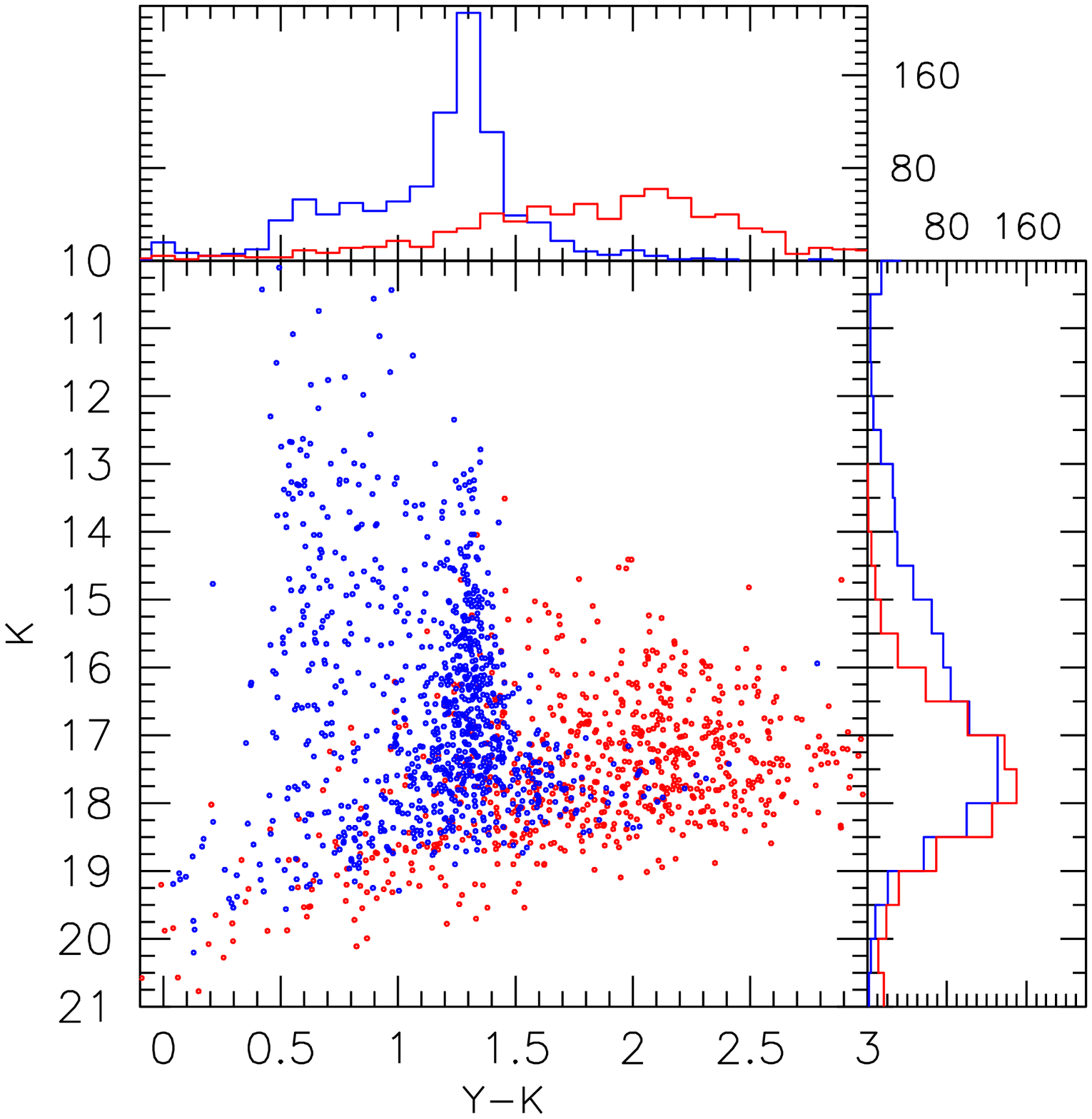}}
\end{minipage}
\caption{A comparison between the aperture photometry 
from UKIDSS image data (left) and the corresponding simulation
(right), for a 0.21 $\sqdeg$ area towards $\ell=-220\degr,
b=40\degr$. The photometry was performed using both DAOPHOT and
SExtractor. The main panels show the CMD obtained combining DAOPHOT
photometry with SExtractor star/galaxy classification (blue/red dots,
respectively). The histograms show the total colour and magnitude
distributions of stars and galaxies (blue and red lines,
respectively).}
\label{fig_ukidss}
\end{figure*}

We have taken the original image from the UKIDSS archive, and
performed aperture photometry with both DAOPHOT \citep{daophot} ad
SExtractor \citep{SExtractor}. A image for the same area has been
simulated using UKIDSS specifications (pixel scale, SNR, etc.) and
submitted to the same kind of catalogue
extraction. Fig.~\ref{fig_ukidss} shows the results, comparing the
$K$ vs. $\yk$ diagrams for the UKIDSS observed (left panel) and simulated
(right panel) fields, for both stars (blue points) and galaxies (red
points). Stars and galaxies were separated using the SExtractor
Stellarity parameter $st$. We adopted $st>0.85$ for stars, and
$st\le0.85$ for galaxies.

Both DAOPHOT ad SExtractor aperture photometries turned out to be
remarkably consistent with the ones provided by the Cambridge
Astronomical Survey Unit (CASU) data reduction pipeline. This is very
comforting since the CASU will adopt the same data reduction pipeline
to the future VISTA data. The histograms at the right and top of the
CMD panels show the object number count distribution in both colour
and magnitude. As can be appreciated, our simulated objects distribute
very similarly in colour and magnitude as the observed ones. The
discrepancies are limited to a few aspects of the simulations, for
instance the peak in the colour distribution at $\yk\sim 1.3$ is
clearly narrower in the models than in the simulations. This peak is
caused by thin disk dwarfs less massive than $0.4\,M_\odot$
\citep{Marigo_etal03}, and its narrowness in the models could be
indicating that TRILEGAL underestimates the colour spread of these
very-low mass stars.

The most important point of the model--data comparison of
Fig.~\ref{fig_ukidss}, however, is that the simulations reasonably
reproduce the numbers (with errors limited to $\sim20$\%), magnitudes
and colours of the observed objects. This gives us confidence that our
MC simulations contain the correct contribution from foreground Milky
Way stars and background galaxies.

%%%%%%%%%%%%%%%%%%%%%%%%%%%%%%%%%%%%%%%%%%%%%%%%%%%%%%%%%%%
\section{Recovering the SFH}
\label{sec_sfh}

\subsection{Basics}

The basic assumption behind any method to recover the SFH from a
composite stellar population (CSP) is that it 
can be considered as a simply sum of its constituent parts,
which are ultimately simple stellar populations (SSPs) or 
combinations of them.
Therefore determining the SFH of any CSP -- like the field stars in a
galaxy -- means to recover the relative weight of each SSP.
The modern stellar population analysis in the
late 80's \citep{TGF89, Ferraro_etal89} and in the 
early 90's \citep{Tosi_etal91, Bertelli_etal92} -- marked by the 
advent of the first CCD detectors -- was done assuming 
parameterized SFH, which revealed the main trends in the SFH but was
still limited by a small number of possible solutions.
The techniques to recover the SFH from a resolved stellar population 
started to become more sophisticated 
with the works of \citet{Gallart_etal96a, Gallart_etal96c}, but they 
were significantly improved by \citet{Aparicio_etal97} and 
\citet{Dolphin97}, who developed for the first time statistical 
methods to recover non-parameterized SFH from the CMD of a CSP. In
practice these two works were the first to deal with a finite
number of free independent components, obtained by adding the
properties of SSPs inside small, but finite, age and metallicity
bins. These ``partial models'' \citep{Aparicio_etal97} are 
so computed for age and metallicity bins that should be small enough
so that the SSP properties change just little inside them, and large
enough so that the limited number of bins ensures reasonable CPU times
for the SFH-recovery. Furthermore, being the partial models
computed for the same and constant star formation rate inside each age
bin, implied that they needed to be generated only once, saving a large
amount of computational resources \citep{Dolphin02}.
% In computing the partial models, a constant star formation rate 
% is assumed inside each age bin.

Considering that a CMD is a distribution of points in a plane divided
into $N_{\rm box}$ boxes, these ideas can be expressed
\citep{Dolphin02} by
\begin{equation}
m_i = \sum_j r_j c_{i,j}
\end{equation}
where $m_i$ is the number of stars in the full model CMD for an CSP in
the $i^{\rm th}$ box, $r_j$ is the SFR for the $j^{\rm th}$ partial
model, and $c_{i,j}$ is the number of stars in the CMD for the $j^{\rm
th}$ partial model in the $i^{\rm th}$ CMD box.

%%%%%%%%Figure %%%%
%\begin{figure}
%\resizebox{\hsize}{!}{\includegraphics{hess_modelling_lowres.ps}}
%\caption{Example of a $(K, \yk)$ Hess diagrams for a LMC stellar population 
%in our simulations: \citet{Marigo_etal08} isochrones constraining the
%stellar ages ($9.00 \leq\log(t/{\rm yr})\leq9.20$) and metallicities
%($Z=0.008$ or $\mh=-0.37$) (panel a); distribution of stars in
%accordance to the IMF, $f_{\rm bin}$=0.30 (panel b) and after the
%effect of photometric errors and completeness (panel c); final Hess
%diagram with colour-coded logarithm of the density of stars (panel
%d). }
%\label{hess_modelling}
%\end{figure}

\begin{figure*}
\begin{minipage}{0.85\textwidth}
\resizebox{\hsize}{!}{\includegraphics{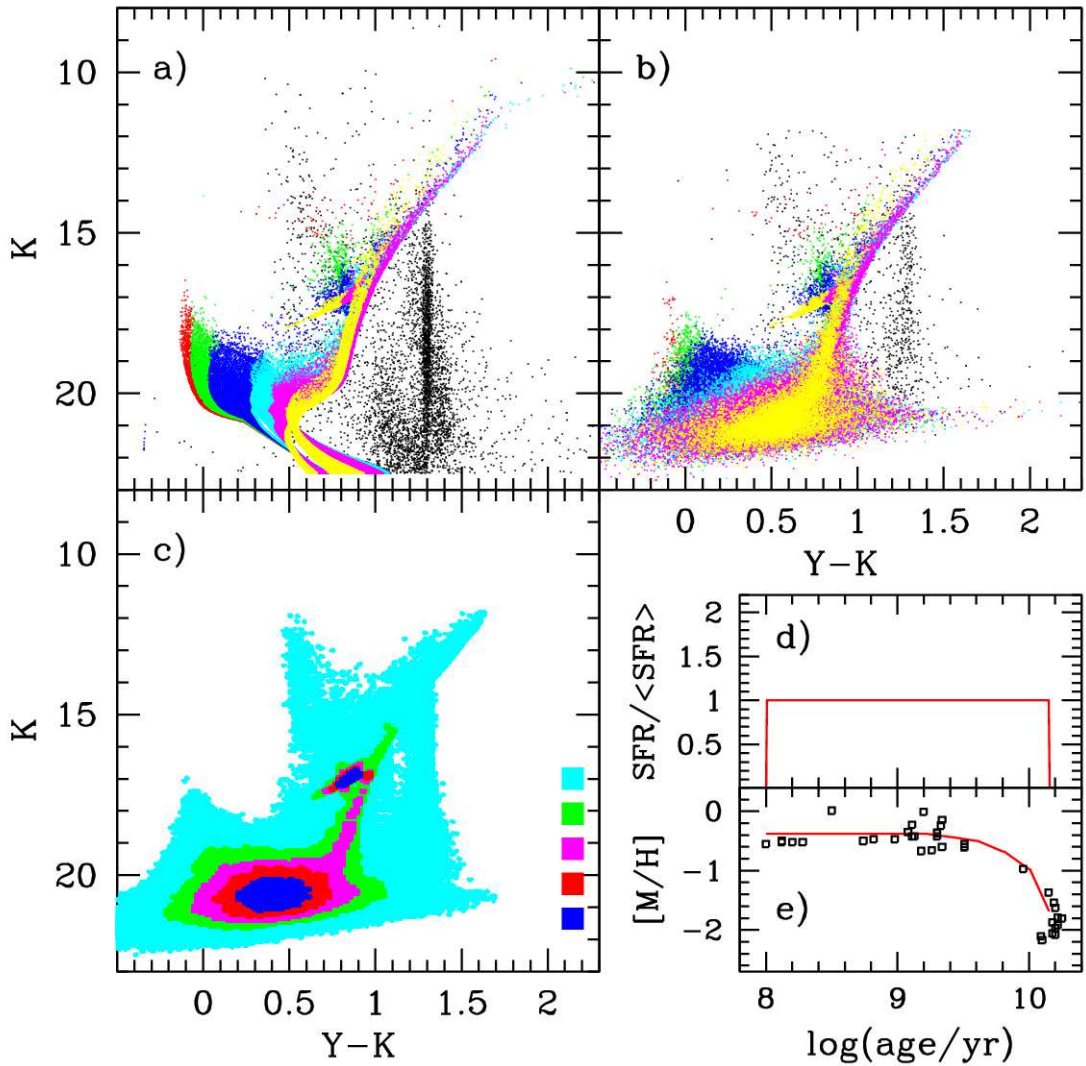}}
\end{minipage}
\caption{Simulated $(K, \yk)$ CMD illustrating the building of 
partial models for the analysis of LMC stellar populations and its
foreground MW stars. Panel (a) shows the theoretical stars generated
from TRILEGAL, corresponds to the following ranges in $\log(t/{\rm
yr})$: 8.00--8.40 (red), 8.40--8.80 (green), 8.80--9.20 (blue),
9.20--9.60 (cyan), 9.60--10.00 (magenta), 10.00--10.15 (yellow), plus
the foreground MW (black). Panel (b) shows the same after considering
the effects of photometric errors and completeness. Panel (c) is the
Hess diagram for the sum of all partial models. Panels (d) and (e)
show the input SFR$(t)$ and AMR, respectively, the latter in
comparison with LMC clusters \citep[squares -- data from][]{MG03,
Kerber_etal07, Grocholski_etal06, Grocholski_etal07}.  To avoid an
extremely large size for this figure only 5\% of all stars typically
used ($\sim10^7$) to build the partial models are shown in the top
panels. }
\label{cmd_starpops}
\end{figure*}

Notice that the above equation is in fact written in terms of Hess
diagrams since we are dealing with the number of stars in CMDs.  So it
means that the ``observed'' Hess diagram for a CSP can be described as
a sum of independent synthetic Hess diagrams of partial models,
being the coefficients $r_{\rm j}$ the SFRs to be determined.
%This provides an efficient use of all information
%available not only in magnitude {\textit or} colour, 
%but in the plane formed by them.  
Fig.~\ref{cmd_starpops}, to be commented later, illustrate the 
generation of such synthetic Hess diagrams for the partial models 
of the LMC.

The classical approach to determine the set of $r_{\rm j}$s is to
compute the differences in the number of stars in each CMD box between
data and model, searching for a {\it minimisation of a
chi-squared-like statistics}. This kind of approach 
was applied for the first time to recover the SFH of a real 
galaxy by \citet{Aparicio_etal97}, and has been
successfully used in the analysis of the field stars in the dwarf
galaxies in the Local Group \citep{Gallart_etal99,
Dolphin02, Dolphin_etal03, Skillman_etal03, Cole_etal07, YL07},
including the MCs \citep{Olsen99, Holtzman_etal99, HZ01,
Smecker-Hane_etal02, HZ04, Javiel_etal05, CV07, Noel_etal07}. 
Although these works have in common the same basic idea of how 
to recover the SFH, there are clear variations in terms of the 
adopted statistics and strategy to divide the CMD in boxes 
\citep{Gallart_etal05}.
   
An interesting alternative to recover the SFH from the analysis of
CMDs is offered by the {\it maximum likelihood} technique 
using a Bayesian approach
\citep{TS96, Hernandez_etal99, Hernandez_etal00, Vergely_etal02}. 
In this approach the basic idea is to establish for each observed star
a probability to belong to a SSP (based on the expected number of
stars from this SSP in the position of the observed star in the
CMD). Doing it for all observed stars, one can recover the SFRs that
maximises the likelihood between data and model.  It is interesting to
note that in the recent years there is an increasing number of papers
applying this kind of technique for a wide range of problems, which
include the determination of physical parameters of stellar clusters
\citep{JL05, NJ06, HVG08} as well as of individual stars
\citep{Nordstrom_etal04, daSilva_etal06}.

It is beyond the scope of the present work to discuss in depth the
particularity of each aforementioned approach, but there are no strong
reasons to believe that one can intrinsically recover a more reliable
SFH than the other \citep{Dolphin02, Gallart_etal05}. So for a
question of simplicity and coherence with the majority of the works
devoted to the MCs, we adopted the classical {\it minimisation of a
chi-squared-like statistics} technique to determine the expected
errors in the SFH for the VMC data, using the framework of the
StarFISH code \citep{HZ01, HZ04}, the $\chi^{2}$-like
statistics defined by \cite{Dolphin02} 
assuming that stars into CMD boxes follow a Poisson-distributed data,
and a uniform grid of boxes in the CMD.

\subsection{StarFISH and TRILEGAL working together}

The StarFISH code\footnote{Available at
http://www.noao.edu/staff/jharris/SFH/} has been developed by
\citet{HZ01} and successfully applied by
\citet{HZ04} and \citet{Harris07} to recover SFHs for the MCs inside the context of 
the Magellanic Clouds Photometric Survey
\citep[MCPS]{Zaritsky_etal97}\footnote{http://ngala.as.arizona.edu/dennis/mcsurvey.html}. This code, originally designed to analyse CMDs built with $UBVI$ data 
from the MCPS, and using Padova isochrones \citep{Girardi_etal00,
Girardi_etal02}, offers the possibility of different choices for
generating synthetic Hess diagrams (set of partial models, CMD binning
and masks, combination of more than one CMD, etc.) and $\chi^{2}$-like
statistics, being also sufficiently generic to be implemented for new
stellar evolutionary models, photometric systems, etc.

\begin{figure*}
\begin{minipage}{0.70\textwidth}
\begin{minipage}{0.30\textwidth}
\fbox{\resizebox{\hsize}{2.00\hsize}{\includegraphics{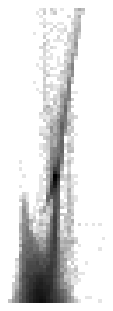}}}
\end{minipage}
\hfill
\begin{minipage}{0.30\textwidth}
\fbox{\resizebox{\hsize}{2.00\hsize}{\includegraphics{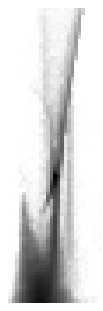}}}
\end{minipage}
\hfill
\begin{minipage}{0.30\textwidth}
\fbox{\resizebox{\hsize}{2.00\hsize}{\includegraphics{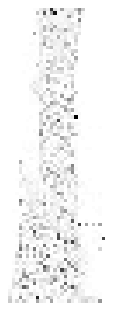}}}
\end{minipage}
\end{minipage}
\hfill
\begin{minipage}{0.27\textwidth}
\caption{
Example of simulated $(K, \yk)$ Hess diagrams for the VMC (left panel)
and for the best solution found by StarFISH (middle panel). The
$\chi^{2}$-like statistics map is also shown in the right panel. These Hess diagrams
are limited to $12.00 < K < 20.50$ and $-0.50 < \yk < 2.20$ and were
built with bin sizes of 0.10~mag both in colour and in magnitude. }
\label{hess_res}
\end{minipage}
\end{figure*}

As illustrated in 
%Figs. \ref{hess_modelling} and \ref{cmd_starpops},
Fig. \ref{cmd_starpops},
the TRILEGAL code can also simulate the synthetic Hess diagram for a
set of partial models, with the advantage of easily generating them in
the UKIDSS photometric system, allowing also a greater control of all
input parameters involved (see Sect.~\ref{lmc_stars}). 
Therefore we decide to directly provide these Hess diagrams to StarFISH, 
using it as a platform to determine the SFRs for our VMC simulated data by
means of a $\chi^{2}$-like statistics minimisation. 
The search of the best solution is
done internally in StarFISH by the {\it amoeba} algorithm that uses a
downhill strategy to find the minimum $\chi^{2}$-like statistics value.

%Fig.~\ref{hess_modelling} illustrates the process of generating the
%synthetic ($K, \yk$) CMDs and Hess diagrams for a single partial
%model, comprising all stars in a specific age interval and for a given
%metallicity. It shows how the stars produced by TRILEGAL are
%distributed in the CMD, before and after the simulation of photometric
%errors and incompleteness, and finally in a colour-coded Hess
%diagram. The figure shows clearly that the simulated stars are not
%limited to CMD areas between the limiting isochrones, first because of
%the presence of detached binaries (panel b), and second because of
%photometric errors at the faintest magnitudes (panel c).

An extra possibility
%Another possibility 
offered by TRILEGAL is the construction of an
additional partial model for the MW foreground\,\footnote{See
http://stev.oapd.inaf.it/trilegal}. Indeed, this is done by simulating
the MW population towards the galactic coordinates under examination,
for the same total observed area but averaging over many simulations
so as to reduce the Poisson noise. This partial model is provided to
StarFISH and used in the $\chi^{2}$-like statistics minimization 
together with those
used to describe the MC population. With this procedure, the presence
of the MW foreground is taken into account in the SFH determination,
without appealing to the (often risky) procedures of {\em statistical
decontamination} based on the observation of external control
fields. To our knowledgement, this is the first time that such a
procedure is adopted in SFH-recovery work. Notice that, once the MW
foreground model is well calibrated, its corresponding $r_{\rm j}$
could be set to a fixed value, instead of being included into the
$\chi^{2}$-like statistics minimization.

Figure~\ref{cmd_starpops} illustrates the generation of a complete set
of partial models, 
covering ages from $\log(t/{\rm yr})=8.00$ to $10.15$ 
($t$ from $0.10$ to $14.13$~Gyr) divided into 11 elements with a 
width of $\Delta\log t=0.20$ each, and following
an AMR consistent with the LMC clusters (see the panel d, and
Sect.~\ref{lmc_stars}). In this figure we have grouped the partial
models in just 6 age ranges (plus the MW foreground one) just for a
question of clarity. What is remarkable in the figure is the high
degree of superposition of the different partial models over the RGB
region of the CMD -- except of course for the partial model
corresponding to the MW foreground. The MS region of the CMD, instead,
allows a good visual separation of the different populations over the
entire age range, even after considering the effects of photometric
errors and incompleteness.

\subsection{Results: Input vs. output SFR$(t)$ }

An example of SFH-recovery is presented in the Hess diagrams of
Fig.~\ref{hess_res}. The input simulation (left panel) was generated
for a constant SFR$(t)$, for an area equivalent to 1 VIRCAM detector
($0.037~\sqdeg$) inside a region with a stellar density typical for
the LMC disk ($\log N_{\rm RGB}=2.00$, that produces a total number of
stars of $N_{\rm stars}\sim 5\times10^{4}$). For this simulation,
StarFISH fits the solution represented in the middle panel. Not
surprisingly, the data--model $\chi^2$-like statistics 
residuals (right panel) are remarkably evenly distributed across the 
Hess diagram.

\begin{figure}
\resizebox{\hsize}{!}{\includegraphics{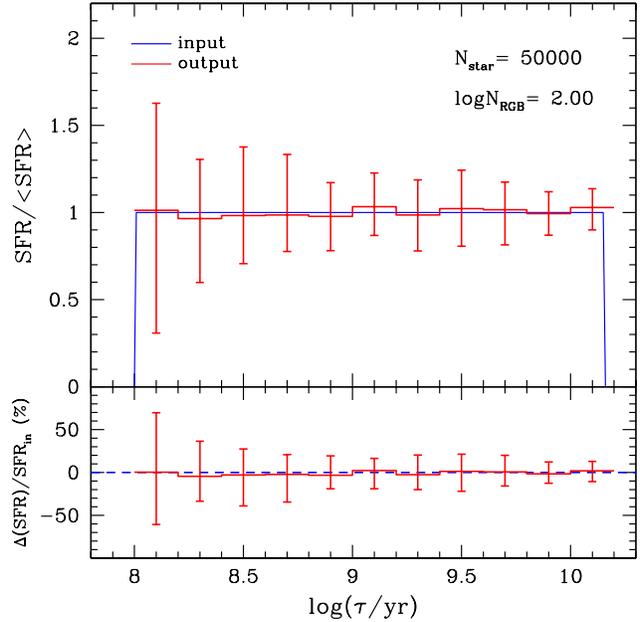}}
\caption{
Errors in the recovered SFR$(t)$ in terms of the mean SFR$(t)$ (top
panel) and input SFR$(t)$ (bottom panel).  The input simulations
correspond to a typical LMC disk region ($\log N_{\rm RGB}=2.00$)
inside a single VIRCAM detector ($\sim 0.037$~$\sqdeg$). The central
solid line corresponds to the median solution found over 100
realizations of the same simulation, whereas the error bars correspond
to a confidence level of 70\%.
%The other red lines outline the same confidence level for an decreasing 
%statistics in the number of stars ($\sim$ 200, 100, 50 $\times10^{3}$ stars)
%or covered area (4, 2, 1 VIRCAM detectors $\sim 0.15, 0.08, 0.04$ $\sqdeg$)
}
\label{error_sfr}
\end{figure}

Figure~\ref{error_sfr} presents the median of recovered SFR$(t)$, and
its error, obtained after performing 100 realizations of the same
simulation. As expected, the median SFR$(t)$ over this 100
realizations reproduces remarkably well the input one, with no
indication of systematic errors in the process of
SFH-recovery\footnote{As a consequence, the total integral of
the SFR is also well recovered.} . The error bars correspond to a
confidence level of 70\%, which means that 70\% of all individual
realizations are confined within these error bars. Error bars are
almost symmetrical with respect to the expected SFR(t).  
Furthermore, errors are typically below 0.4 in units of mean
SFR$(t)$ (top panel), which means uncertainties below 40\% (bottom
panel) for almost all ages. The only exception is the youngest age bin
which presents errors in the SFR that are about two times larger
than those for the intermediate-age stellar populations.

There are a many factors that can affect the accuracy of a recovered
SFR$(t)$. Among them, the most important are:
\begin{enumerate}
\item
the quality of the data in terms of stellar statistics and photometry,
that depends in principle (for the same photometric conditions of
seeing, exposure times, calibration, etc.) on the density of the field
and its covered area in the sky;
\item
the uncertainties in the models themselves, that come from the
possible errors in the adopted input parameters (distance, reddening,
IMF, $f_{\rm bin}$, AMR, etc.), in the stellar evolutionary models,
and in the imperfect reproduction of photometric errors and
completeness;
\item
the incorrect representation of the contamination from other sources,
like foreground MW stars, stars from LMC star clusters, and background
galaxies;
\item
the non-uniform properties of the analysed field, like differential 
reddening or depth in distance.
\end{enumerate}
Notice that the first factor affects the generation of observed Hess
diagrams, while the synthetic Hess diagrams may become unrealistic due
to the other factors. Furthermore, they produce different types of
errors: whereas the first preferentially rules the {\it random}
errors, the second is the main source of the {\it systematic}
errors. A discussion on errors is addressed below.

\subsubsection{Random errors for a known AMR}

\begin{figure}
\resizebox{\hsize}{!}{\includegraphics{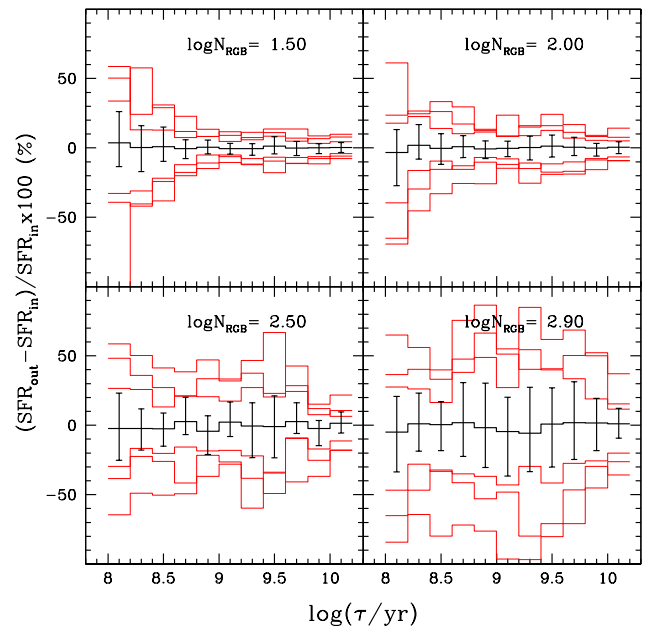}}
\caption{Errors in the recovered SFR$(t)$ 
for four different stellar densities, from the outer LMC disk 
(top-left panel) to the LMC centre (bottom-right panel).  
The thick black solid line corresponds to the median solution found 
using $4\times10^{5}$ stars, whereas the error bars correspond to a 
confidence level of 70\%.
%The other red lines outline the same confidence level for an decreasing 
%statistics in the number of stars ($\sim$ 200, 100, 50 $\times10^{3}$ stars)
%or covered area (4, 2, 1 VIRCAM detectors $\sim 0.15, 0.08, 0.04$ $\sqdeg$)
The thin red lines outline the same confidence level for a decreasing
number of stars: $2\times10^{5}$, $10^{5}$, $5\times10^{4}$ stars.  }
\label{error_sfr_vs_dens}
\end{figure}

In order to estimate the expected random errors in the SFR$(t)$ for
the LMC we performed controlled experiments similar to the one shown
in Fig.~\ref{error_sfr}, but covering a wide range of conditions in
regard to the stellar statistics and
crowding. Fig.~\ref{error_sfr_vs_dens} presents the results for four
different levels of density, from the outer LMC disk (top-left panel)
to the LMC centre (bottom-right panel), for a number of stars (or
area) varying by a factor of 8. As can be seen, these two factors
dramatically change the level of accuracy than can be achieved in the
recovered SFR$(t)$: an increase in the number of stars reduces the
errors whereas an increase in crowding for a fixed number of stars
acts in the opposite way.

\begin{figure}
\resizebox{\hsize}{!}{\includegraphics{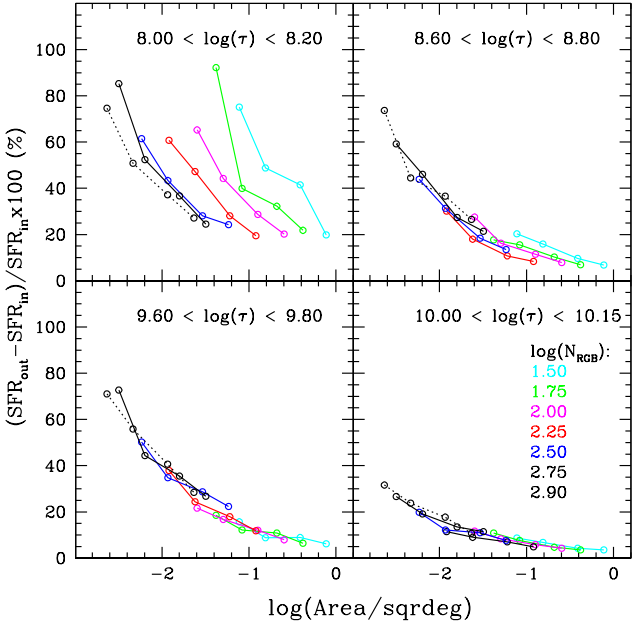}}
\caption{Errors in the recovered SFR$(t)$ as a function of
covered area and density of stars (different lines) for 4 partial
models of different ages (different panels).}
\label{error_sfr_vs_area}
\end{figure}

The errors in the recovered SFR$(t)$ as a function of the covered
area, for all simulated density levels, are shown in
Fig.~\ref{error_sfr_vs_area} for partial models of four different
ages, from young (top-left panel) to old ones (bottom-right
panel). Here, a very interesting result can be seen: for stars older
than $\log(t/{\rm yr})\sim 8.60$ ($t\sim 0.4$ Gyr) the curves for
different levels of density are almost superposed, revealing that for
a fixed area the accuracy in the recovered SFR$(t)$ is roughly
independent of the level of crowding.  It can be understood as a
counterbalanced effect between the loss of stars due to a decrease in
completeness, and the gain of stars due to an increase in
density. Therefore, it seems that the SFR$(t)$ for this wide range in
age can be determined with random errors below 20\% if an area of
$0.10\,\sqdeg$ is used. Increasing the area by a factor of four means
that the level of uncertainty drops to below 10\%.

On the other hand, for partial models younger than $\log(t/{\rm
yr})\sim 8.60$ ($t\sim 0.4$ Gyr) a fixed area does not imply a
constant level of accuracy in the SFR$(t)$, being the errors
significantly greater in the less dense regions. Since stars in this
small age range are mainly identified in the upper main sequence at
$18\la K\la20$, and in the core-helium burning phases at $14\la
K\la18$ (see Fig.~\ref{cmd_starpops}), this effect can be understood
by the fact that for these brighter stars the increase in stellar
density is not followed by a significant decrease in
completeness. Indeed, Fig.~\ref{photo_all} indicates the
completeness is in general higher than 60\% and 90\%, for $18\la
K\la20$ and $14\la K\la18$ stars, respectively, for the entire range
of stellar densities of the LMC. In this situation, what determines
the accuracy for a fixed area is simply the number of observed stars,
which is proportional to the density. In particular, our simulations
reveal the lack of stellar statistics in the outermost and less dense
LMC regions, which require great areas to reach a statistically
significant number of young stars. For instance, a level of
uncertainty 20\,\% in the recovered young SFR$(t)$ is obtained for an
area $\sim10$ times larger ($\sim 1\,\sqdeg$) in the periphery of the
LMC than in more central regions.

\begin{figure}
\resizebox{\hsize}{!}{\includegraphics{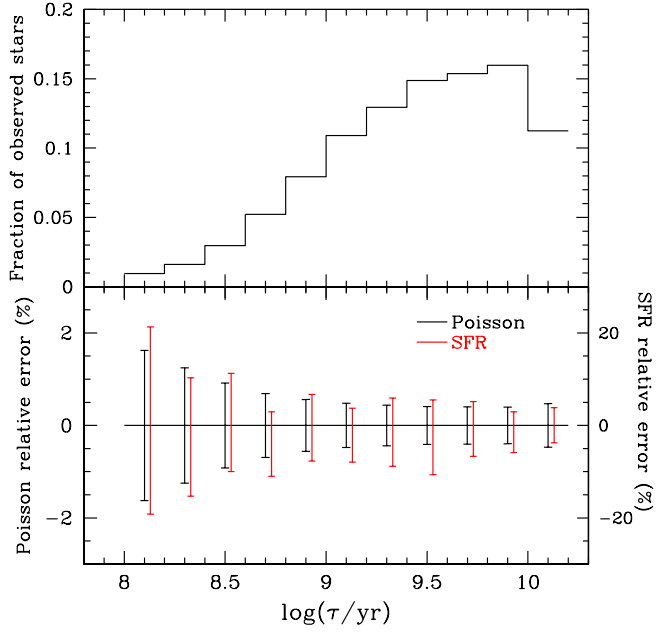}}
\caption{Distribution of the fraction of observed stars (top panel)
and its Poisson relative errors (bottom panel, black lines) as a
function age.  These errors correspond to $4\times10^{5}$ simulated
stars. The relative errors in the recovered SFR$(t)$ are also shown (bottom
panel, red lines) for a vertical scale 10 times larger (compare the
labels of the two vertical axis in the bottom panel). }
\label{poisson_errors}
\end{figure}

The errors in the SFR$(t)$ were also compared with those expected by
the Poisson statistics in the number of observed stars for each
partial model, as presented by Fig.~\ref{poisson_errors}. This figure
reveals that the errors in the SFR$(t)$ can be roughly understood as a
propagation of the Poisson errors by a factor $\sim 10$. This explains
why errors are smaller for the older ages, despite the fact that the
different partial models are -- due to the adoption of a logarithmic
age scale -- roughly uniformly separated in the CMD.

%\subsection{Comparison with the accuracy of previous works }

\subsubsection{Random errors for an unknown AMR}

\begin{figure}
\resizebox{\hsize}{!}{\includegraphics{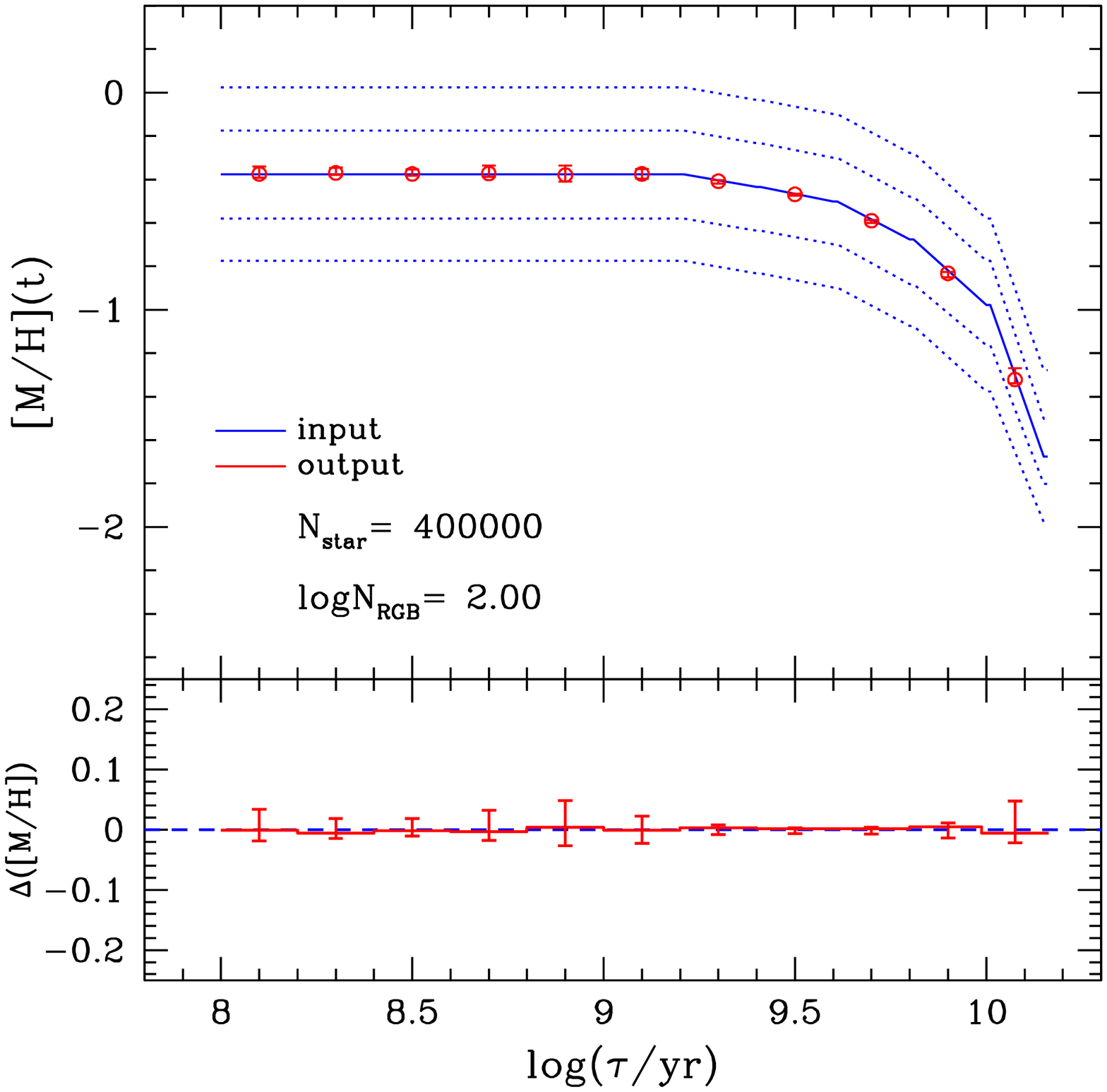}}
\caption{The top panel shows the distribution of metallicities of the 
partial models adopted in this work: The central solid line is the AMR
adopted as a reference, and is used in all of our SFH-recovery
experiments. At every age (or age bin), 4 additional partial models
(along the dotted lines) can be defined and inserted in the
SFH-recovery, then allowing us to access the AMR and its uncertainty
(see text for details). The bottom panel shows the difference between
the input and output AMRs for the case in which 5 partial models were
adopted for each age bin. }
\label{error_MH_400000_2.00}
\end{figure}

\begin{figure*}
\begin{minipage}{0.49\textwidth}
\resizebox{\hsize}{!}{\includegraphics{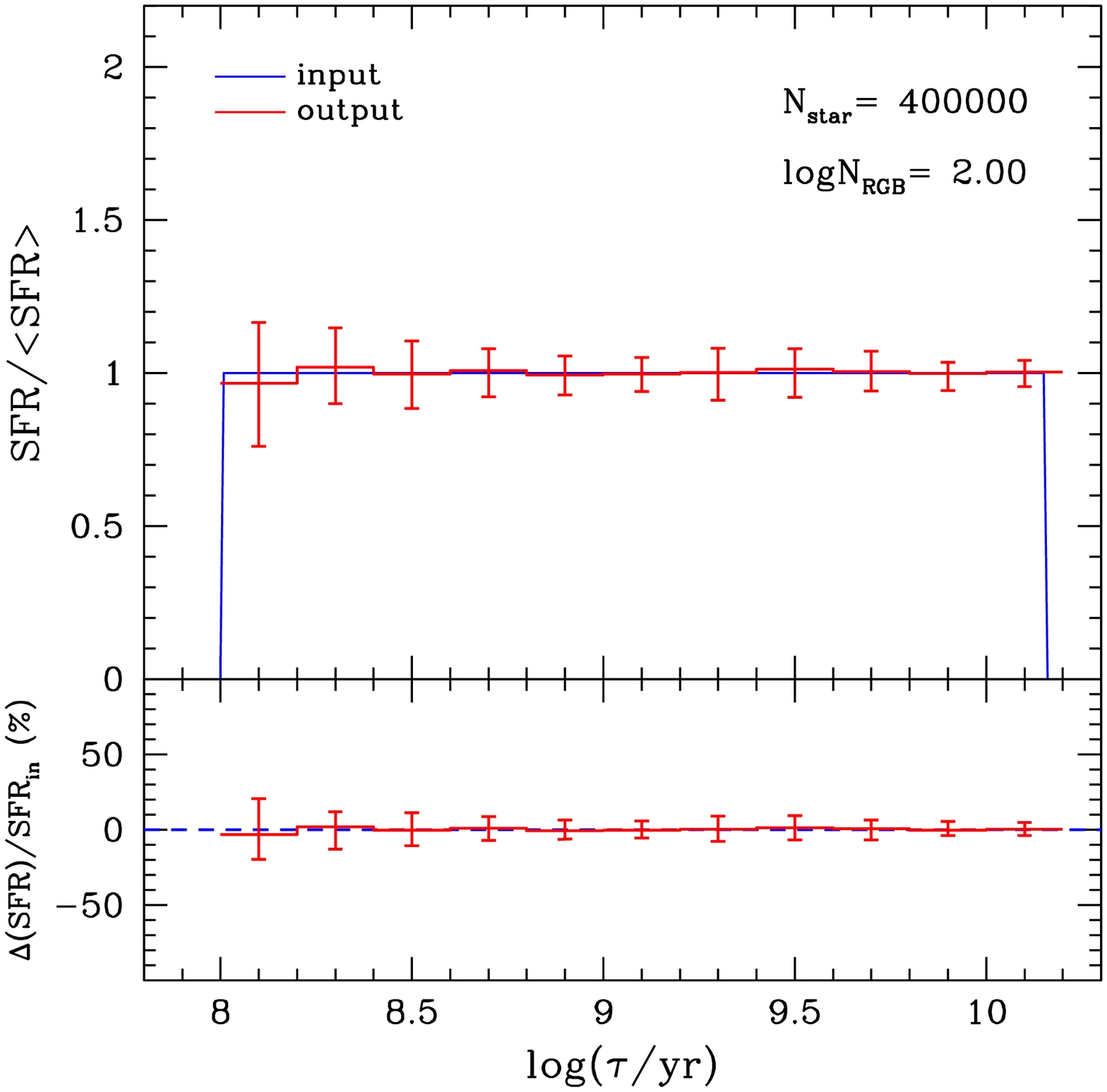}}
\end{minipage}
\hfill
\begin{minipage}{0.49\textwidth}
\resizebox{\hsize}{!}{\includegraphics{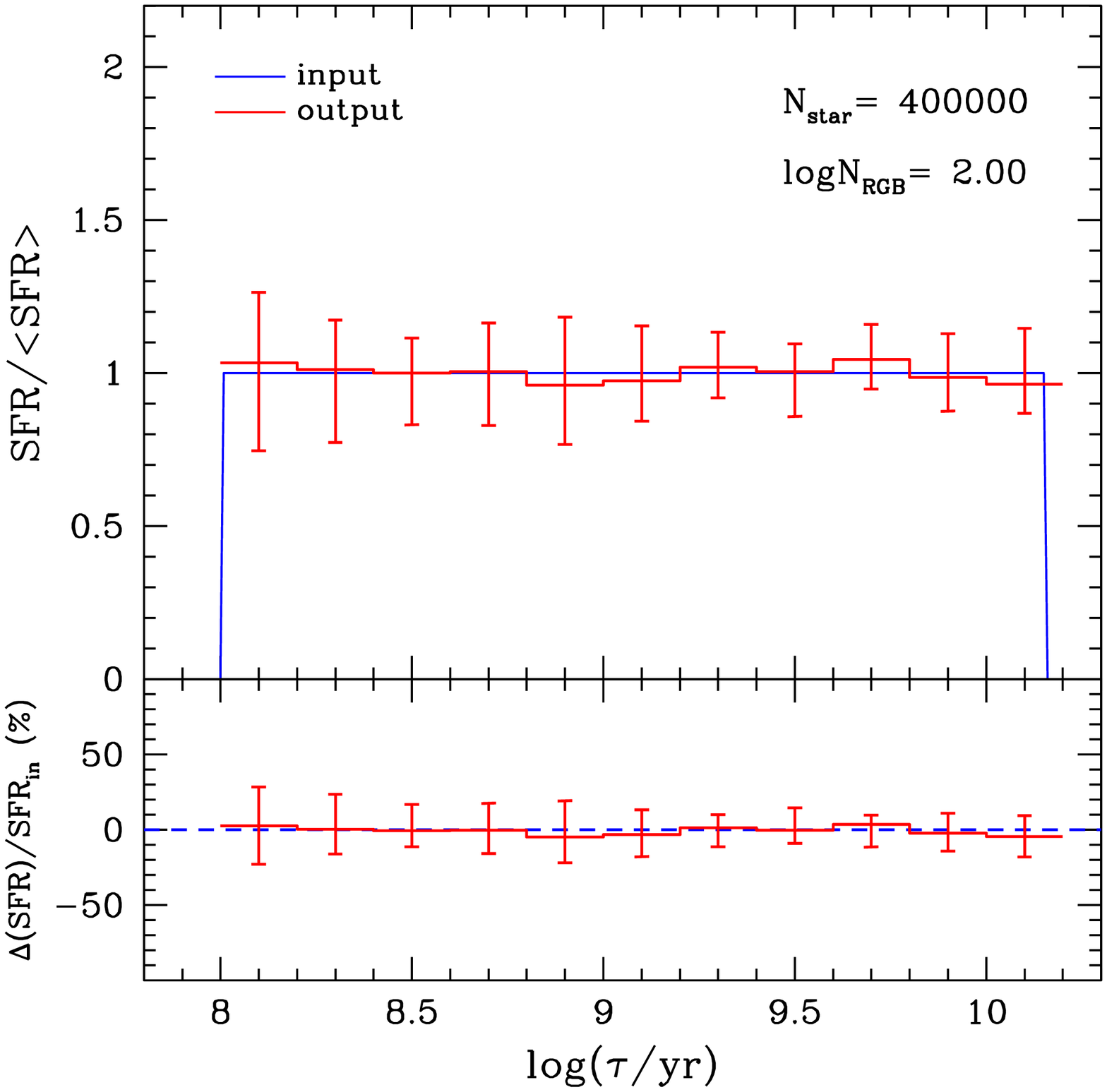}}
\end{minipage}
\caption{Errors in the recovered SFR$(t)$ in terms of the mean 
SFR$(t)$ (top panels) and input SFR$(t)$ (bottom panels), for a
typical LMC disk region inside the area of 8 VIRCAM detector
($0.25\,\sqdeg$). The left panel correspond to a SFH-recovery that
uses partial models distributed over a single AMR, whereas the right
one uses 5 partial models for each age bin.  The central solid line
corresponds to the median solution found over 100 realizations of the
same simulation, whereas the error bars correspond to a confidence
level of 70\%.}
\label{error_sfr_400000_2.00}
\end{figure*}

In our previous discussion we have naively assumed that the AMR of the
LMC is well known, by using a set of partial models which strictly
follows the AMR used in the simulations. The real situation is much
more complicated. Not only the AMR is not well established, but also
it may present significant spreads (for a single age), and vary from
place to place over the LMC disk. In order to face this situation, it
is advisable to allow a more flexible approach for the SFH-recovery,
in which for every age bin we have different partial models covering a
significant range in metallicity.

We adopt the scheme illustrated in the top panel of
Fig.~\ref{error_MH_400000_2.00}, that is: for every age bin, we build
partial models of 5 different metallicities, centered at the \mh\
value given by the reference AMR, and separated by steps of
0.2~dex. This gives a total of 56 partial models, and drastically
increases the CPU time (by a factor of $\sim100$) needed by StarFISH
to converge towards the $\chi^2$-like statistics minimum. 
Once the minimum is found,
for each age bin we compute the $r_j(t)$-added SFR$(t)$
and the $r_j(t)$-weighted average $\mh(t)$ over the five 
partial models with different levels of metallicity.
%, and the %$r_j(t)$-weighted average for the AMR $\mh(t)$. 
After doing the same for 100 realizations of the input simulation, 
we derive the median and the confidence level of 70\% 
(assumed as the error bar) for the output SFR$(t)$ and $\mh(t)$.
Figs.~\ref{error_MH_400000_2.00} and
\ref{error_sfr_400000_2.00} illustrate the results for the case of a
constant input SFR$(t)$ and $\log N_{\rm RGB}=2.00$.

In the top panel of Fig.~\ref{error_MH_400000_2.00}, the dots with
error bars show the output AMR, which falls remarkably close to the
input one (continuous line). The error bars are smaller than 0.1~dex
at all ages.  The bottom panel plots the relative errors in the
derived $\mh(t)$, showing that they are slightly larger for
populations of age $\log(t/{\rm yr})<9.2$. Anyway, the main result
here is that the errors in $\mh(t)$ are always smaller than the
0.2~dex separation between the different partial models.

Figure~\ref{error_sfr_400000_2.00} instead compares, for the same
simulation, the errors in the SFR$(t)$ that result either considering
(right panel) or not (left panel) the partial models with metallicity
different from the reference AMR one. In other words, the left panel
shows the SFR$(t)$ that would be recovered if the AMR were exactly
known in advance, whereas the right panel shows the SFR$(t)$ for the
cases in which the AMR is unknown -- or, alternatively, affected by
significant observational errors. It can be easily noticed that the
SFR$(t)$ is correctly recovered in both cases, although errors in the
second case (right panel) are about 2 or 3 times larger than in the
first case.  Needless to say, the second situation is the more
realistic one, and will likely be the one applied in the analysis of
VMC data.

\subsubsection{Systematic errors related to distance and reddening}

Accessing all the systematic errors in the derived SFR$(t)$ is beyond
the scope of this paper. However, the errors associated with the
variations in the distance and reddening are of particular interest
here, since both quantities are expected to vary sensibly across the
LMC, hence having the potential to affect the patterns in the
spacially-resolved SFH. Fortunately, these errors are very easily
accessed with our method, since we know exactly the $\dmo$ and $A_V$
values of the simulations, as well as those assumed during the
SFH-recovery.

\begin{figure}
\resizebox{\hsize}{!}{\includegraphics{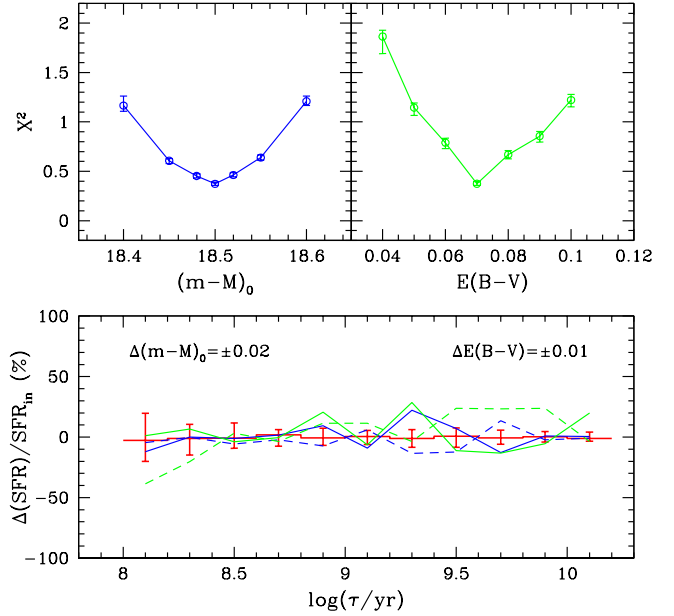}}
\caption{
The top panels show the minimum $\chi^{2}$-like statistics 
value as a function of distance modulus (blue) and reddening (green) 
adopted to build the synthetic Hess diagrams. 
The bottom panel illustrate the systematic
variations that the errors in these two parameters cause in the
recovered SFR$(t)$ (continuous lines for an overestimation of
$(m-M)_0$ and \ebv, dashed lines for an underestimation). }
\label{chi2_vs_parameter}
\end{figure}

The propagation of the uncertainties in the assumed distance modulus
and reddening in the recovered SFR$(t)$ was explored in a series of
control experiments, in which a simulation performed with the
canonical $\dmo=18.50$ and $\ebv=0.07$ was submitted to a series of
SFH-recovery analyses covering a range of \dmo\ and \ebv. More
specifically, \dmo\ was varied from 18.40 to 18.60, and \ebv\ from
0.04 to 0.1. The results of these tests can be seen in
Fig.~\ref{chi2_vs_parameter}, which presents the effect of wrong
choices in these parameters not only in the recovered SFR$(t)$ (bottom
panel) but also in the minimum value for the 
$\chi^{2}$-like statistics (top panels).

As expected, the absolute minimum value for the 
$\chi^{2}$-like statistics was found
for the synthetic Hess diagrams with the right distance modulus and
reddening.  Based on the $\chi^{2}$-like statistics dispersion 
for 100 simulations we estimated that the errors in these parameters are about
$\Delta\dmo=\pm 0.02$~mag and $\Delta\ebv=\pm0.01$~mag. These errors
also imply systematic errors in the recovered SFR$(t)$ of up to
$\sim30$\% (Fig.~\ref{chi2_vs_parameter}, bottom panel).

The above experiments clearly suggest us that the mean distance
and reddening should be considered as free parameters in the analysis,
and varied by a few 0.01~mag so that we can identify the best-fitting
values of \dmo\ and \ebv\ together with the best-fitting SFH. This
kind of procedure has been adopted by e.g. \citet{Holtzman_etal97} and
\citet{Olsen99} in their study of small regions over the LMC, and it
is also implemented in the MATCH SFH-recovery package by
\citet{Dolphin02}. Occasionally, one could also consider small spreads
in both \dmo\ and \ebv\ and test whether they further improve the
$\chi^2$-like statistics minimization. 
Once applied to the entire VMC area, the final
result of this procedure will be independent maps of the geometry and
reddening across the MC system, that can complement those obtained
with other methods.

%%%%%%%%%%%%%%%%%%%%%%%%%%%%%%%%%%%%%%%%%%%%%%%%%%%%%%%%%%%
\section{Concluding remarks}
\label{sec_conclu}

In this paper we have performed detailed simulations of the LMC
images expected from the VMC survey, and analysed them in terms of the
expected accuracy in determining the space-resolved SFH. Our main
conclusions so far are the following:
\begin{enumerate}
\item For a typical $0.10~\sqdeg$ LMC field of median stellar density, the 
random errors in the recovered SFR$(t)$ will be typically smaller than
20\% for 0.2~dex-wide age bins. 
\item For all ages larger than 0.4~Gyr, at increasing stellar densities the 
better statistics largely compensates the effects of increased
photometric errors and decreased completeness, so that good-quality
SFR$(t)$ can be determined even for the most crowded regions in the
LMC bar. The SFR$(t)$ errors decrease roughly in proportion with the
square of the total number of stars. The exception to this rule
regards the youngest stars, which because of their brightness are
less affected by incompleteness. In this latter case, however, the
stellar statistics is intrinsically small and large areas are
necessary to reach the same SFR$(t)$ accuracy as for the
intermediate-age and old LMC stars.
\item Although the AMR [M/H]$(t)$ can be recovered with accuracies better 
than 0.2~dex, the uncertainties in the AMR can significantly affect
the quality of the derived SFR$(t)$, increasing their errors by 
a factor of about 2.5.
\item The minimisation algorithms allow to identify the best-fitting 
reddening and distance with accuracies of the order of 0.02~mag in
distance modulus, and 0.01~mag in \ebv.
\end{enumerate}

All of the above trends were derived from analysis of small LMC areas,
that we have considered to be homogeneous in all of their properties
(AMR, distance and reddening). The errors were derived by varying each
one of these parameters separately. The real situation will be, of
course, much more complicated, with significant spatial variations of
all of these quantities across the LMC. This consideration may lead us
to suppose that errors here derived are underestimated. However, the
above-mentioned parameters can be further constrained by simply taking
into consideration additional data -- for instance the available
reddening maps, the limits on the relative distances provided by other
independent distance indicators, the metallicity distributions of
field stars, etc. -- in our analysis. Moreover, our work indicates
clearly how the random errors are reduced when we increase the area to
be analysed. It is natural that, once the systematic errors are fully
assessed, we will increase the area selected for the analysis, so that
random errors become at least smaller than the systematic ones.

It is also worth mentioning that our present results were obtained
using the $\yk$ colour only. VMC will also provide CMDs involving the
$J$ passband, and their use in the SFH analysis can only reduce the
final errors.

Another factor to be considered, in the final analysis, is that for
old ages the SFR$(t)$ is expected to vary very smoothly across the
LMC, as indicated for instance by \citet{Cioni_etal00} and
\citet{NW00}. This large-scale correlation in the old SFR$(t)$
may be used as an additional constraint during the SFH-recovery, and
may help to reduce the errors in the SFR$(t)$ at all ages.

A forthcoming paper will discuss in more detail the expected accuracy
over the complete VMC area, including the SMC, and how this accuracy
depends on other variables and constraints which were not discussed in
this work (binaries, depth in distance, reddening variations, IMF,
etc.).  Anyway, the present work already illustrates the excellent
accuracy in the measurements of the space-resolved SFH, that will be
made possible by VMC data. Moreover, it demonstrates that detailed
SFH-recovery using deep near-infrared photometry is also feasible, as
much it has always been for the case of visual observations.

%%%%%%%%%%%%%%%%%%%%%%%%%%%%%%%%%%%%%%%%%%%%%%%%%%%%%%%%%%%%%%%%%%%%%%%%%%
\begin{acknowledgements}
This work is based in part on data obtained as part of the UKIRT
Infrared Deep Sky Survey. This publication makes use of data products
from the Two Micron All Sky Survey, which is a joint project of the
University of Massachusetts and the Infrared Processing and Analysis
Center/California Institute of Technology, funded by the National
Aeronautics and Space Administration and the National Science
Foundation.

We acknowledge the referee for useful comments and suggestions, 
that helped us to significantly improve the quality of the paper, 
specially concerning the discussion about the different methods to 
recover the SFH (Sect.~\ref{sec_sfh}).
We are very grateful to Jim Emerson for providing preliminary VISTA
filter throughputs. L.G. acknowledges the
inspiring role of J. Dalcanton, A. Dolphin, J. Harris, J. Holtzman,
K. Olsen, and B. Williams (and their papers) in some of the aspects
implemented/tested in this work. We acknowledge support from the
Brazilian funding agency CNPq, and from INAF/PRIN07 CRA 1.06.10.03.
\end{acknowledgements}

% for the bibliography, at the end
\bibliographystyle{aa} %style aa.bst
\bibliography{vmcpaper} % your references file.bib

\end{document}